%% file: wlenet.tex
\documentclass[fleqn,usenatbib]{mnras}
\usepackage{newtxtext,newtxmath}

\usepackage[T1]{fontenc}
\usepackage{ae,aecompl}

\usepackage{graphicx}	
\usepackage{amsmath}	
\usepackage{amssymb}	
\usepackage{float}
\usepackage{anyfontsize} 
\usepackage{url}
\usepackage{booktabs}
\usepackage{tabularx}
\usepackage[utf8]{inputenc}

\DeclareMathOperator*{\E}{\mathbb{E}}

\DeclareMathOperator*{\argmin}{arg\,min}

\title[Weak lensing: a machine learning approach]{Weak lensing shear estimation beyond the shape-noise limit: a machine learning approach}

\author[O. M. Springer et al.]{
Ofer M. Springer,$^{1,2}$\thanks{E-mail: springer@phys.huji.ac.il (OMS)}
Eran O. Ofek,$^{2}$
Yair Weiss$^{1}$
and Julian Merten$^{3,4}$
\\
$^{1}$School of Computer Science and Engineering, The Hebrew University of Jerusalem, Jerusalem, Israel\\
$^{2}$Benoziyo Center for Astrophysics, Weizmann Institute of Science, 76100 Rehovot, Israel\\
$^{3}$INAF-OAS Osservatorio di Astrofisica e Scienza dello Spazio di Bologna, Via Gobetti 93/3, 40129, Bologna, Italy\\
$^{4}$Department of Physics, University of Oxford, Denys Wilkinson Building, Keble Road, Oxford OX1 3RH, UK
}

\date{Accepted 2019 October 16. Received 2019 September 23; in original form 2018 August 22}

\pubyear{2018}

\hypersetup{draft}

\begin{document}
\label{firstpage}
\pagerange{\pageref{firstpage}--\pageref{lastpage}}
\maketitle

\begin{abstract}
Weak lensing shear estimation typically results in per galaxy statistical errors significantly larger than the sought after gravitational signal of only a few percent. These statistical errors are mostly a result of \textit{shape-noise} --- an estimation error due to the diverse (and a-priori unknown) morphology of individual background galaxies. These errors are inversely proportional to the limiting angular resolution at which localized objects, such as galaxy clusters, can be probed with weak lensing shear. In this work we report on our initial attempt to reduce statistical errors in weak lensing shear estimation using a machine learning approach --- training a multi-layered convolutional neural network to directly estimate the shear given an observed background galaxy image. We train, calibrate and evaluate the performance and stability of our estimator using simulated galaxy images designed to mimic the distribution of \textit{HST} observations of lensed background sources in the CLASH galaxy cluster survey. Using the trained estimator, we produce weak lensing shear maps of the cores of 20 galaxy clusters in the CLASH survey, demonstrating an RMS scatter reduced by approximately 26\% when compared to maps produced with a commonly used shape estimator. This is equivalent to a survey speed enhancement of approximately 60\%. However, given the non-transparent nature of the machine learning approach, this result requires further testing and validation. We provide python code to train and test this estimator on both simulated and real galaxy cluster observations. We also provide updated weak lensing catalogues for the 20 CLASH galaxy clusters studied.
\end{abstract}



\section{Introduction}

In weak lensing, the small gravitational deflection of light in its path from the source to the observer enables the measurement of the intervening mass distribution. Prime examples in which weak lensing is applied are in the measurement of cosmic shear, where the large-scale structure is of interest, and in measurements of galaxy cluster lensing, where the mass distribution is comparably more localized \citep[see][for a review of weak lensing applications]{schneider2006weak}. 

This weak deflection of light induces a small distortion that maps the source plane coordinates $\vec\theta'$ to the image plane coordinates $\vec\theta$. The distortion $\vec\theta'(\theta)$ may be locally approximated as an affine transformation with a Jacobian

\begin{equation}
\label{eq:shear}
\frac{\partial \vec\theta'}{\partial \vec\theta} = 
\begin{pmatrix}
    1-\kappa-\gamma_1 &-\gamma_2 \\
    -\gamma_2 &1-\kappa+\gamma_1
\end{pmatrix} 
= (1-\kappa)
\begin{pmatrix}
    1-g_1 &-g_2 \\
    -g_2 &1+g_1
\end{pmatrix},
\end{equation}
where $\kappa$ is known as the \textit{convergence} and parametrizes the uniform stretching of the image, $\gamma_i$ are the two components of \textit{shear} and $g_i = \gamma_i/(1-\kappa)$ are the $\textit{reduced shear}$. Weak lensing shear estimation is typically concerned with measuring the reduced shear in images of faint, high-redshift galaxies and with deducing the lensing mass distribution from these estimates \citep[see][for a derivation of weak lensing formalism]{bartelmann2001weak}.\\

An ideal estimator of $g_i$ is unable to access the pre-lens galaxy image and therefore has to rely on the (true) probability distribution $P(g_i \lvert x)$ of the shear given galaxy image $x$. Here $x$ represents galaxy image pixel responses as observed, including the effects of galaxy population morphology, lens distortion, point-spread-function (PSF), band throughputs and noise --- only to name a few known observational factors.

To date, many shear estimation techniques have been developed and studied and these have been primarily focused on reducing various types of systematic errors (e.g. biases due to PSF correction, noise, or source population variation). For a recent review of shear estimation methods and the different sources of estimation bias see \cite{mandelbaum2015great3} and \cite{massey2012origins}. \\

Less attention has been given to the issue of reducing the statistical errors of the shear estimator that result from the wide variation in galaxy shapes.  In cluster lensing, the number of background sources is often very limited. The angular resolution of galaxy cluster weak lensing maps is typically limited by the density of available background sources (on which shear can be estimated) and such maps are often computed by locally averaging the shear estimates of only a few neighbouring galaxies at each point in the field. For a given set of background sources, reducing the statistical error (or RMS scatter) of the shear estimator would therefore directly translate to a linear enhancement of the angular resolution of the map (at a pre-selected level of map noise).

In galaxy cluster lensing studies, two shear estimation methods that have been widely used are the \textsc{KSB} method \citep{kaiser1994method} for ground based observations and the \textsc{RRG} method \citep{rhodes2000weak} for space based observations. These estimators essentially measure the two component ellipticity $e$ of the galaxy intensity profile from its weighted quadrupole moments, correcting for the weight profile and PSF in different ways. After calibrating a proper \textit{shear susceptibility factor} $G$ \citep[see][]{rhodes2000weak,leauthaud2007weak}, the calibrated ellipticity $\hat{\epsilon} = e/G$ is used as an unbiased local estimate of the reduced shear $g = \langle \hat{\epsilon} \rangle$. Note that although different definitions of ellipticity exist in the literature, here $\hat{\epsilon}$ has been scaled to match the scale of the reduced shear $g$ defined in equation \ref{eq:shear}. The intrinsic per-component scatter $\sigma_{\hat{\epsilon}}$ of $\hat{\epsilon}$ (at constant $g$), resulting from the natural diversity of galaxy intensity profiles (and not from various measurement errors), is known as \textit{shape-noise} and is generally considered to be a lower-limit of the statistical error that quadrupole moment based shear estimators can reach. The actual value of the shape-noise $\sigma_{\hat{\epsilon}}$ is somewhat dependent on source galaxy population \citep[see][where it is found to be ${\sim}0.25$ and ${\sim}0.30$ at the lower and upper magnitude cuts of a specific space-based survey, respectively]{leauthaud2007weak}.\\

In this work we aim to measure shear with a statistical error lower than $\sigma_{\hat{\epsilon}}$ by using additional morphological information, visible in the source galaxy stamp, beyond its quadrupole moments. The premise of our work is that the apparent structure of galaxies isn't fully described by their ellipticity parameters and that quadrupole-based estimators may therefore be suboptimal with respect to statistical error. Shear estimators that measure higher order intensity moments (above quadrupole) and attempt to model their joint statistics \citep[see e.g.][]{refregier2003shapelets,bernstein2016accurate} could also potentially utilize substructure to achieve this goal, although, to the best of our knowledge, these estimators have not yet demonstrated to do so. 

The idea that there is additional information to be harnessed for weak lensing measurements beyond the imaged quadrupole moments has been suggested before. Additional observables such as radio polarization were used by \cite{brown2011mapping} to constrain the intrinsic morphological orientation of galaxies. Resolved spectroscopic kinematic maps were used by \cite{blain2002detecting,morales2006technique,huff2013cosmic} and more recently \cite{de2015direct}, to reduce the shape noise in measurements of disk galaxies. Dimensionality reduction techniques of existing multi-band photometric data, when used in addition to quadrupole moment ellipticities, have also been shown to carry some additional information \citep{croft2017prediction,niemi2015weak}. In contrast, in this work, we use only a single optical image as an observable and seek new image features that may be used to infer shear.

Our approach to this problem is based on a discriminative machine learning model --- a multi-layered (or \textit{deep}) \textit{convolutional neural network} trained to directly estimate shear (and not ellipticity) from the observed galaxy stamps. This approach relies on the availability of a large dataset of simulated galaxy stamps, having known shear and simulated observational conditions matching those of the galaxy clusters under study. We stress that this requirement is both challenging and crucial for properly training and validating a shear estimator in such a machine learning approach.\\

This paper is structured as follows: In \S\ref{sec:observations} we describe the \textit{Hubble Space Telescope} (\textit{HST}) observations, both of the survey in which we measure shear and of the survey from which we generate simulated training and calibration data for the estimator. In \S\ref{sec:simulations} we describe the steps taken to match the simulations to the galaxy cluster observations and to generate a larger and richer training dataset that would promote certain useful invariance properties of the learnt estimator. In \S\ref{sec:model} we describe the machine learning model used and its training procedure. In \S\ref{sec:performance_sim} we calibrate biases and measure the model performance using simulated data. In \S\ref{sec:clash_analysis} we use the learnt model to measure shear in real images of galaxy clusters, also assessing bias (relative to RRG estimates) and statistical performance on the cluster data. We describe our code and data release in \S\ref{sec:reproducibility} and conclude in \S\ref{sec:discussion}.


\section{Observations}
\label{sec:observations}

In this section we describe the \textit{HST} observations and input catalogues used in this work. These served both to produce simulations, based on real galaxy images drawn from COSMOS observations \citep{scoville2007cosmos,koekemoer2007cosmos}, and to perform weak lensing analysis of the CLASH galaxy cluster data \citep{postman2012cluster}. We also describe here the additional photometric measurements and cuts performed to supplement the input catalogues. These were needed to enable the matching of the COSMOS and CLASH distributions in the simulation phase discussed in the following section.

\subsection{CLASH data}

The premise of our work is that galaxy shape estimation methods may not fully capture the lensing shear signal in an observational regime where a large fraction of the lensed sources are well resolved. We therefore focus our attention on space based observations, specifically the galaxy cluster core observations available in CLASH \citep{postman2012cluster}. The full CLASH data set\footnote{\url{https://archive.stsci.edu/prepds/clash/}} provides \textit{HST} observations in 16 bands for 25 galaxy clusters, enabling also a robust estimation of photometric redshifts for the detected background sources \citep[produced by the CLASH collaboration using the \textsc{BPZ} algorithm of][]{benitez2000bayesian}. Of these 25 clusters, five could not be included in our analysis, either due to a missing F625W filter image (Abell 611, MACSJ0744+39, MACSJ1423+24), missing distortion parameters (CLJ1226+3332) or a guide star astrometric failure (MACSJ0717+37). The 20 clusters we do include are listed in Table~\ref{tbl:clash_catalog_wl}. 

In this work, the CLASH observations serve a dual purpose. First, we would like our simulations to be grounded to a specific observational setting. Second, we would like to demonstrate that our estimator, which is trained and tested on simulated examples, is also able to estimate shear on real galaxy cluster observations, producing results which are at least consistent with those obtained from existing shear estimation methods. 

\input{clash_clusters.tex}

The RRG \citep{rhodes2000weak} shape catalogue produced for the weak lensing analysis of the CLASH \textit{HST} observations by \cite{merten2015clash} is the baseline we choose to compare our method to. The RRG method was specifically designed to reduce shape measurement errors for well resolved galaxies and is commonly used in weak lensing analyses of \textit{HST} observations. Using the full CLASH catalogue, the shape catalogue of \cite{merten2015clash} was produced by first selecting background sources using \textsc{BPZ} photometric redshifts. PSF corrected RRG shapes were measured for each background source on $0.03\arcsec$ pixel-scale co-added images. This was done per filter, per visit, allowing dense \textsc{TinyTim} PSF models to be estimated in the field of view and reference frame of each co-added image \citep{krist201120,rhodes2007stability}. These shape measurements were then rotated to a J2000 north-up reference frame and combined per background source by computing \textit{signal-to-noise} (S/N) weighted averages.

\subsubsection{ACS total images}
\label{sec:acs_total_images}

For our own weak lensing analysis of the CLASH data, we choose to estimate shear on the same set of filters used by \cite{merten2015clash} but using the $0.065\arcsec$ pixel-scale, north-up, \textsc{MosaicDrizzle} images made available by the CLASH collaboration. Although this pixel-scale is considered sub-optimal for weak lensing analysis \citep[see][]{rhodes2007stability}, this choice allows us to reduce the effects of noise correlations in the simulations as discussed in \S\ref{sec:simulations}. To simplify the current analysis and to maximize S/N, we also choose to measure shear in each cluster on a single total filter-co-added image. We define this image and its accompanying inverse variance image below.

Although these two choices may introduce additional systematic and statistical errors (e.g. co-addition of images having different PSF) in our weak lensing analysis of the CLASH clusters (\S\ref{sec:clash_analysis}), they do not affect the validity of the performance metrics evaluated on the simulations (\S\ref{sec:performance_sim}). Moreover, in our CLASH analysis, after calibrating a linear bias model relative to the baseline RRG shears, we demonstrate that our results are both consistent with those of the baseline and show a relative reduction in statistical errors comparable to the one we measure on simulations. In \S\ref{sec:discussion} we discuss ways in which these design choices can be avoided in future studies. 

\begin{figure}
\begin{center}
\includegraphics[width=1.0\columnwidth]{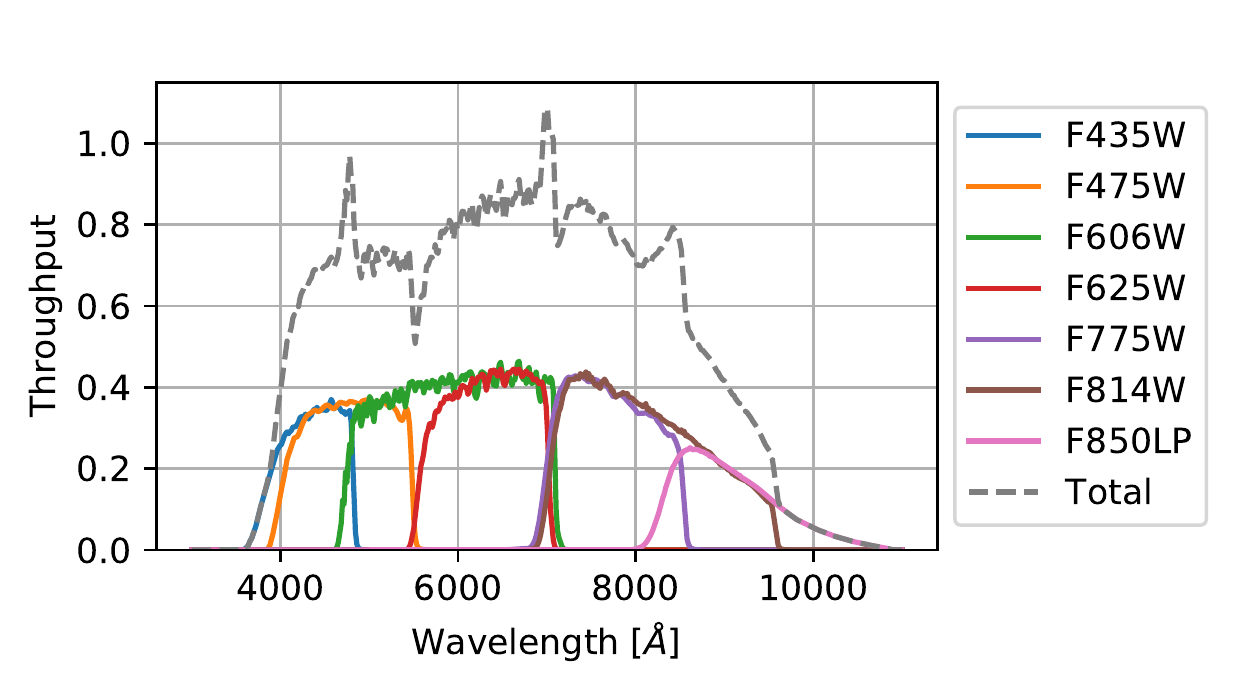}
\end{center}
\vspace*{-3mm}
\caption{Throughputs of the 7 ACS optical filters used in this work. We also plot the throughput of the effective filter resulting from their addition in the \textit{ACS total} images.}
\label{fig:clash_throughputs}
\end{figure}

We now describe how these \textit{total images} were computed. Starting with the available optical filter exposures (see Figure~\ref{fig:clash_throughputs}) taken by the \textit{Advanced Camera for Surveys} \citep[\textit{ACS};][]{ford2003overview} on-board the \textit{HST}, CLASH performed image reduction, alignment, co-addition and produced per filter images using the \textsc{MosaicDrizzle} pipeline at an output pixel-scale of $0.065\arcsec$. Each such filter image is accompanied by an inverse variance image which encodes at each pixel the noise from accumulated dark current, detector readout, and photon noise from the background, but not the additional Poisson noise from the astronomical sources in the image \citep[see][\S5.8.1]{koekemoer2011candels}. We denote the set of $n=7$ filter images by $I_i$ and the inverse variance images by $V_i^{-1}$. The total images and total inverse variance images that are used in this work (for each cluster field) are then simply 
\begin{eqnarray}
I_\textrm{total} &=& \frac{1}{n}\sum_i I_i,\\
V_\textrm{total}^{-1} &=& n^2\left(\sum_i V_i\right)^{-1}.
\end{eqnarray}
 We choose to use these total images because they have a spatially constant and well defined spectral throughput (see \textit{total} in Figure~\ref{fig:clash_throughputs}), as opposed to the \textit{inverse variance weighted} total images made available by the CLASH collaboration \citep{postman2012cluster}.

\begin{figure}
\includegraphics[width=\columnwidth]{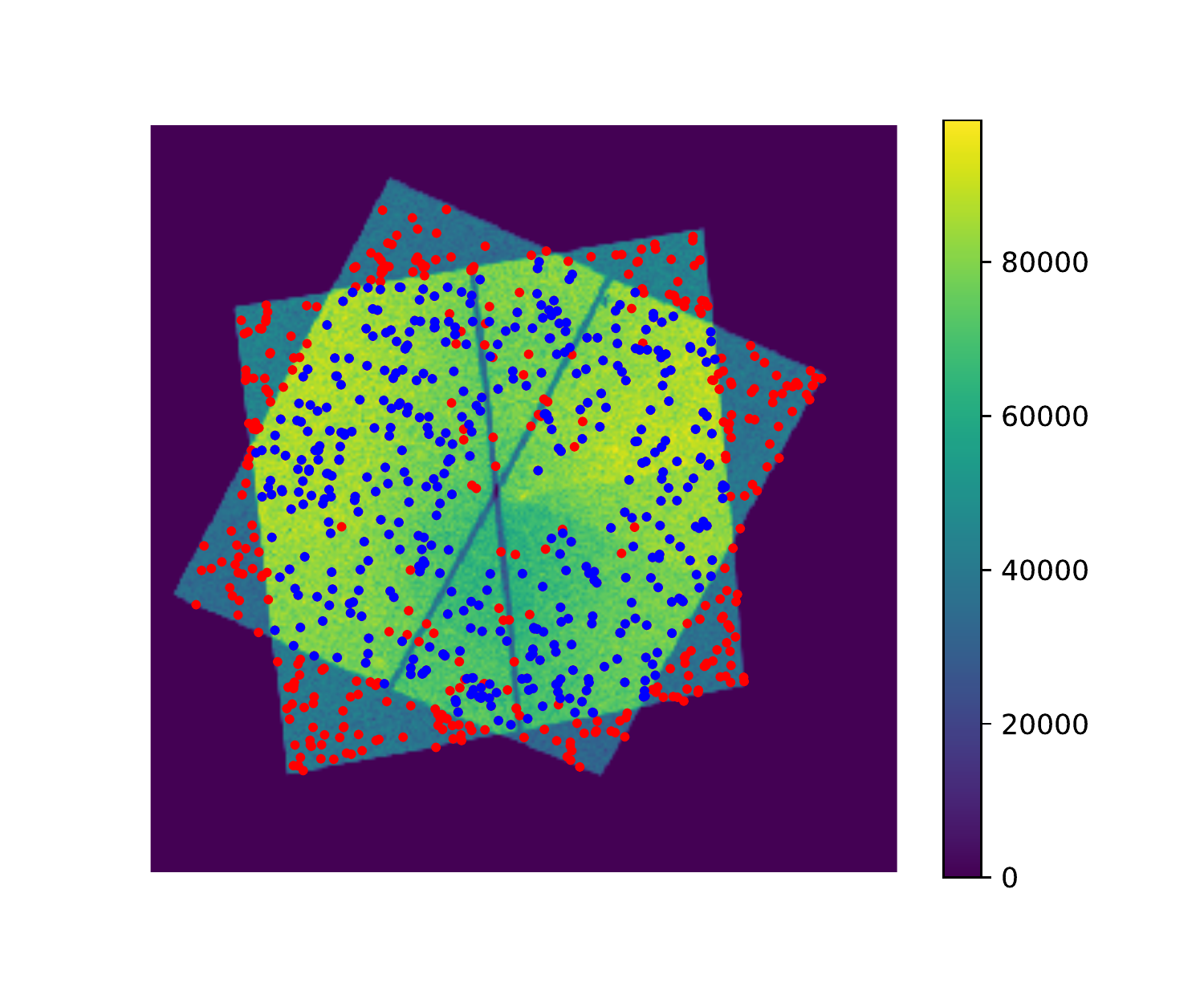}
\vspace*{-5mm}
\caption{Inverse variance weight map for representative CLASH cluster Abell 1423, with weak lensing catalogue overlaid. Blue (red) points mark locations of the original (excluded) background sources. A minimum mean weight threshold of $5\times 10^4 \left[\textrm{ADU}^{-2}\right]$ was chosen to exclude underexposed sources.}
\label{fig:clash_weightmap}
\end{figure}

\subsubsection{Photometry}

In addition to the per band photometric catalogue of \cite{postman2012cluster} we measure photometry in the aforementioned ACS total images. Using \textsc{SExtractor} version 2.19.5 \citep{bertin1996sextractor}, we detect sources and measure full-widths at half-maximum (FWHM), magnitudes ($M_\mathrm{total}$), and half-light flux-radii ($r_{1/2}$) on \texttt{AUTO} apertures. We use the configuration parameters used by \cite{postman2012cluster}, and the ACS total inverse variance images (defined in \S\ref{sec:acs_total_images}) as weight maps. These measurements are then assigned to the weak lensing catalogue of \cite{merten2015clash} by image coordinate matching of the two catalogues. Approximately $87\%$ of the weak lensing sources, which were originally detected using additional bands \citep[see the ACS+IR detection image used by][]{postman2012cluster}, are matched to sources within $0.65\arcsec$ in the ACS total images in this way. For presentation purposes, we define the zero-points of the ACS total image magnitudes, $M_{\mathrm{total}}$, such that $\left<M_{\mathrm{total}}\right> = \left<M_{\mathrm{F814W}}\right>$ on all matched sources in the weak lensing catalogues, and $M_{\mathrm{F814W}}$ are the isophotal $\mathrm{F814W}$ magnitudes of \cite{postman2012cluster}.

\subsubsection{Point source detection and additional cuts}
\label{sec:point_sources}

\begin{figure}
\includegraphics[width=\columnwidth]{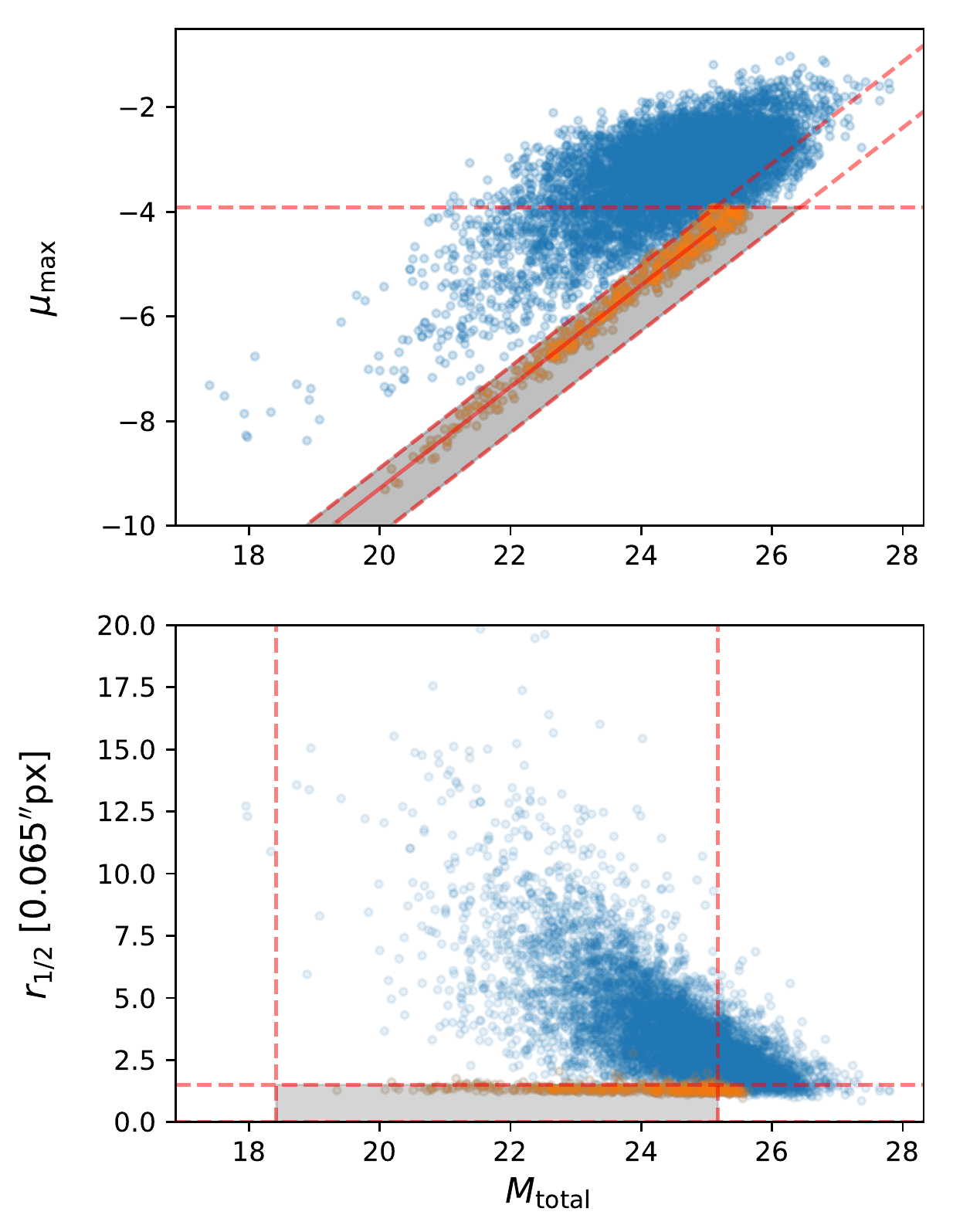}
\vspace*{-7mm}
\caption{Point source detection (orange points) in the combined set of background source catalogues of \protect\cite{merten2015clash}, for the CLASH clusters listed in Table~\ref{tbl:clash_catalog_wl}. Point sources were selected using the procedure of \protect\cite{leauthaud2007weak} and \protect\cite{bardeau2005cfh12k}, by first manually selecting a segment in $r_{1/2}-M_\text{total}$ space (bottom) for which the horizontal branch is well separated. This population was then used to fit a slope and intercept in $\mu_\text{max}-M_\text{total}$ space (top). After also manually selecting a well separating maximal $\mu_\text{max}$ cutoff, the final set of point sources was defined to be those contained in either one of the enclosed regions (grey).}
\label{fig:point_sources}
\end{figure}

To reduce lensing signal dilution due to incorrect inclusion of Milky Way stars in the background source catalogue of \protect\cite{merten2015clash}, we detect point sources using the method described by \cite{leauthaud2007weak} and \cite{bardeau2005cfh12k}. This method makes use of the fact that sufficiently bright stars are usually well separated from galaxies in the radius-magnitude and the $\mu_{\textrm{max}}$-magnitude planes. Here, $\mu_{\textrm{max}}$ is the magnitude of the maximal pixel in the detected aperture of a source. In Figure~\ref{fig:point_sources} we show the distribution of the background sources in these two planes. Sources belonging to either one of the grey regions in Figure~\ref{fig:point_sources} are considered to be stars and excluded from our weak lensing catalogue. \cite{leauthaud2007weak} found that separation using the two planes (radius-magnitude and $\mu_{\textrm{max}}$-magnitude) is overall consistent to about $1\%$ and $2\%$ at \textrm{F814W} magnitudes below $24$ and $25$ respectively, with the difference between the two methods mainly arising from a misclassification of objects in the radius-magnitude plane due to the presence of close pairs. We use the two planes in tandem in the following way. First, the clearly separated horizontal locus is selected in the radius-magnitude plane by selecting lenient upper bounds for the radius and magnitude at which the two populations are well separated (allowing for some point sources to be missed). Using this radius-magnitude selected population of point sources, their locus of points in the $\mu_{\textrm{max}}$-magnitude plane is selected by fitting the intercept and slope in the $\mu_{\textrm{max}}$-magnitude plane and manually setting an upper bound for $\mu_{\textrm{max}}$ (and a horizontal extent for this stripe) at which the populations are well separated. This allows us to use the $\mu_{\textrm{max}}$-magnitude plane for the eventual classification while reducing the number of manually adjusted degrees of freedom.

We measure the median FWHM of these point sources to be $0.121\arcsec$ and use this width (and an isotropic Gaussian profile) as the model PSF in our simulations.

To reduce the variation in the noise properties of the sources in our weak lensing catalogue we additionally exclude sources located in regions of the field that were not fully exposed (see Figure~\ref{fig:clash_weightmap}). We do this by cutting out sources for which the mean value of the inverse variance image in a $32$ pixel wide square aperture is below $5\times 10^4\left[\textrm{ADU}^{-2}\right]$. We also remove sources overlapping the bad-pixel mask of any one of the optical filters.

\begin{figure}
\begin{center}
\includegraphics[width=0.8\columnwidth]{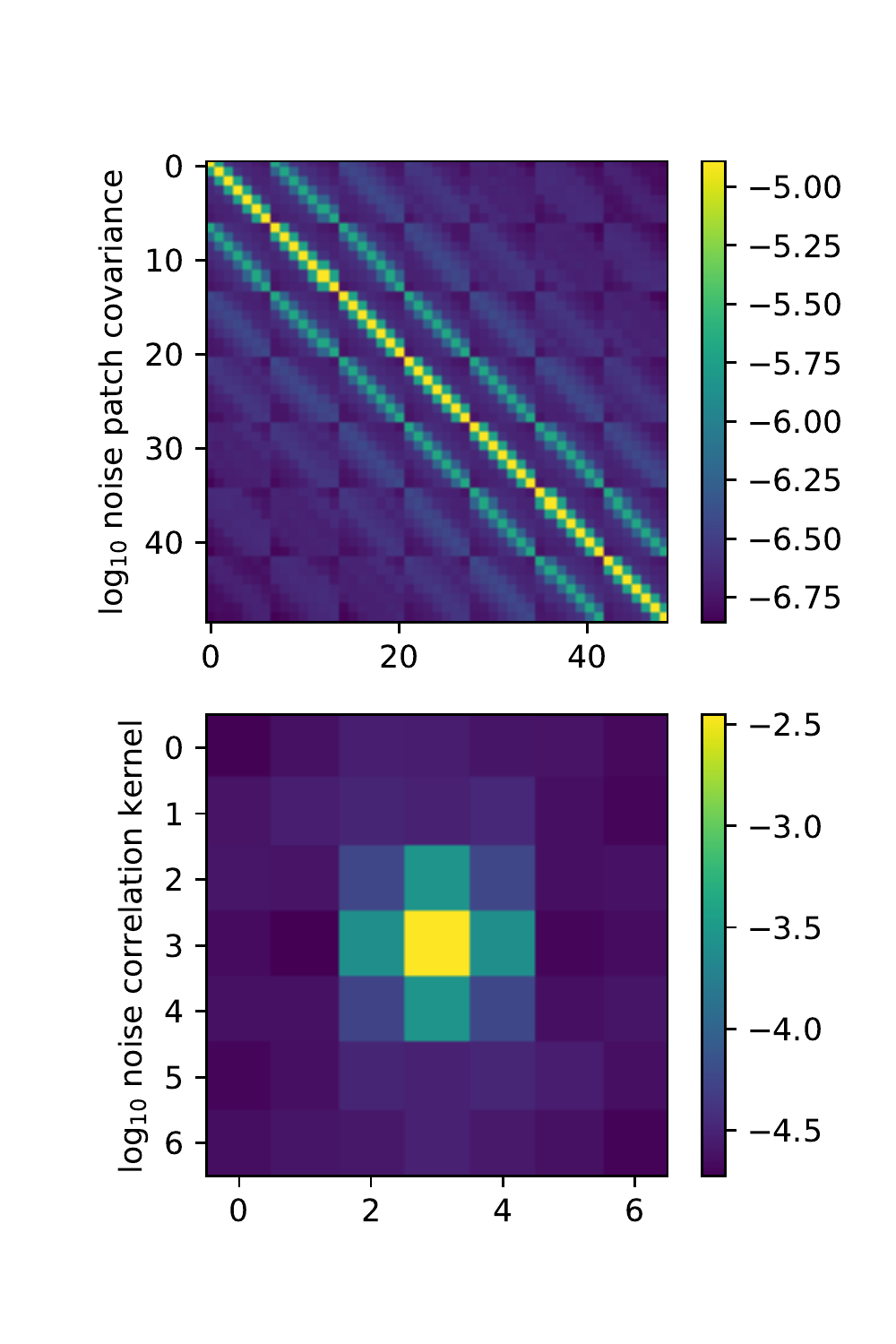}
\end{center}
\vspace*{-6mm}
\caption{Background noise covariance $\Sigma_{\textsc{bg}}$ computed from background subtracted $7\times 7$ pixel patches not overlapping detected sources (top), and noise reproducing kernel $K_{\textsc{bg}}$ (bottom). Convolving $K_{\textsc{bg}}$ with standard normal noise images produces correlated noise having the same local covariance as measured by $\Sigma_{\textsc{bg}}$.}
\label{fig:clash_noise}
\end{figure}

\subsubsection{Correlated background estimation}
\label{sec:correlated_noise}
It is well known that the co-addition of images by the \textsc{MosaicDrizzle} pipeline results in local noise correlations in the output images. To faithfully reproduce the statistics of the CLASH observations in the simulations, we measure the local covariance properties of the background pixels. Using the \texttt{SEGMENTATION} and \texttt{-BACKGROUND} \textit{check images} produced by \textsc{SExtractor} while processing each field, we extract all background subtracted $7\times 7$ pixel patches which do not overlap detected apertures and which survive the same noise exclusion criteria listed in \S\ref{sec:point_sources}. Denoting the $49$-dimensional column vector encoding the $i$'th such patch by $p_i$, then the empirical covariance matrix of $m$ such patches is (see Figure~\ref{fig:clash_noise})

\begin{equation}
\label{eq:sigma_bg}
\Sigma_{\textsc{bg}} = \frac{1}{m-1}\sum_{i=1}^m p_i p_i^{T}.
\end{equation}
Computing a matrix square root of $\Sigma_{\textsc{bg}}$ and extracting the central row of the resulting matrix,

\begin{equation}
K_{\textsc{bg}}  = \left[\sqrt{\Sigma_{\textsc{bg}}}\right]_{i=25},
\end{equation}
we obtain a vector encoding a $7\times 7$ kernel. This kernel is then used in \S\ref{sec:train_and_aug} to generate additive correlated noise having the same covariance properties as the typical CLASH background, by convolving $K_{\textsc{bg}}$ with independent identically distributed (iid) standard normal noise image pixels.

\subsection{COSMOS data}

For the simulations (based on real galaxy images) of the following section we use the two magnitude limited subsets of the \textit{HST} COSMOS survey \citep{scoville2007cosmos,koekemoer2007cosmos}, $M_{\textrm{F814W}} < 23.5,\,25.2$ magnitude, distributed with the \textsc{GalSim}\footnote{\url{https://github.com/GalSim-developers/GalSim}} package \citep{mandelbaum2012precision,rowe2015galsim}. These two datasets consist of random subsets of the full catalogue of \cite{leauthaud2007weak}, containing 56,062 and 87,798 sources respectively, with 24,868 sources contained in their intersection. The datasets include galaxy stamps extracted from registered, undistorted, sky-subtracted images, co-added in an unrotated reference frame at an $0.03\arcsec$ pixel-scale using the \textsc{MultiDrizzle} pipeline \citep{koekemoer2003multidrizzle}. This was done after correcting for the effects of charge transfer inefficiency (CTI) using the method of \cite{massey2009pixel}. Point source detection was performed to remove stars from the source catalogues \citep{leauthaud2007weak,bardeau2005cfh12k}. The datasets additionally include \textsc{TinyTim} PSF models \citep{krist201120,rhodes2007stability} estimated at the extracted source positions as well as a \textsc{SExtractor} \citep{bertin1996sextractor} photometric catalogue, including measurements of F814W AB magnitudes on \texttt{AUTO} apertures and half-light-radii (\texttt{FLUX\_RADIUS} with \texttt{PHOT\_FLUX\_FRAC} set to $0.5$). For a more complete description of the image processing stages, additional catalogue cuts and the photometric measurements performed, the reader is referred to \cite{mandelbaum2012precision} and references therein. 

\section{Simulations}
\label{sec:simulations}

We simulate realistic galaxies using the \textsc{GalSim} package \citep{rowe2015galsim} implementing the \textsc{Reconvolution} algorithm of \cite{mandelbaum2012precision}. \textsc{GalSim} uses the aforementioned COSMOS datasets to render realistic galaxies as they would appear, post shear, in wider PSF observing conditions. Although we required simulations similar to those produced for the realistic, space based, constant shear (RSC) branch of the GREAT3 challenge \citep{mandelbaum2014third}, these were not suitable due to a \textit{shape-noise cancellation} procedure employed there (the inclusion of 90$^\circ$ rotated pairs in each constant shear field, to reduce the effects of shape-noise). Additionally, our aim was to match the distribution of the CLASH observations as closely as possible. We do so with regards to noise properties (\S\ref{sec:sn_tradeoffs}) and the source magnitude-radius distributions (\S\ref{sec:mag_rad_matching}).

\subsection{S/N trade-offs, rescaling and cut-off}
\label{sec:sn_tradeoffs}
In this work, we simulate observations which are intended to match a target population of galaxy images (the CLASH background galaxies) using a source population of galaxy images (the COSMOS in-the-field galaxies). As we discuss in the following section, we choose to match the simulations to the target population by focusing on their distribution in magnitude-radius space. We also require the simulations to reproduce the S/N conditions in the observations. The COSMOS stamps already contain background noise, at a level and having correlations different from the ones we wish to simulate in CLASH. Training our model on simulated galaxies containing this type of correlated noise leads to poor generalization to a validation set and to the real CLASH dataset. Reproducing the S/N conditions was therefore necessary and required particular treatment.

\textsc{GalSim}'s \textit{noise whitening} and \textit{noise symmetrization} features are intended to allow control of the covariance properties of the additive background noise of the simulated target stamps. In the version of \textsc{GalSim} used in this study (version 1.4.4) we found that both \textit{noise whitening} and \textit{noise symmetrization} resulted in correlated noise in the target stamps which depended on the applied shear. Recently \cite{sheldon2017practical} reported difficulties in performing model bias calibration using these features due to what we believe are similar causes.

Unable to reproduce the intended background noise in the stamps at their original scale (using \textsc{GalSim}'s \textit{noise whitening} and \textit{noise symmetrization}), we chose instead to reduce the scale of the stamps and thereby reduce the level of per pixel noise leaking into the simulations. Additionally, in those regions of the stamp where flux from the galaxy was low relative to the noise level (we chose a particular per pixel S/N cutoff of $3.5$), pixel values were nullified. These rescaled and masked COSMOS stamps were then treated as noiseless observations to which the target (pseudo-random) correlated background noise (having the CLASH noise statistics as measured in \S\ref{sec:correlated_noise}) was added.

This procedure introduces a trade-off between an image coordinate rescaling factor $\alpha$, and the extent to which each galaxy aperture is masked (which should both preferably be avoided as much as possible: $\alpha$ should preferably be close to unity to avoid a galaxy age mismatch and the aperature should be as large as possible to avoid a discontinuity at the galaxy outskirts). As discussed above, this trade-off is due to the fact that the target stamp S/N increases with decreasing rescaling factor $\alpha$. In this section we illustrate and quantify this trade-off and report on the particular choices made in our own simulations. \\

\begin{figure}
\begin{center}
\includegraphics[width=\columnwidth]{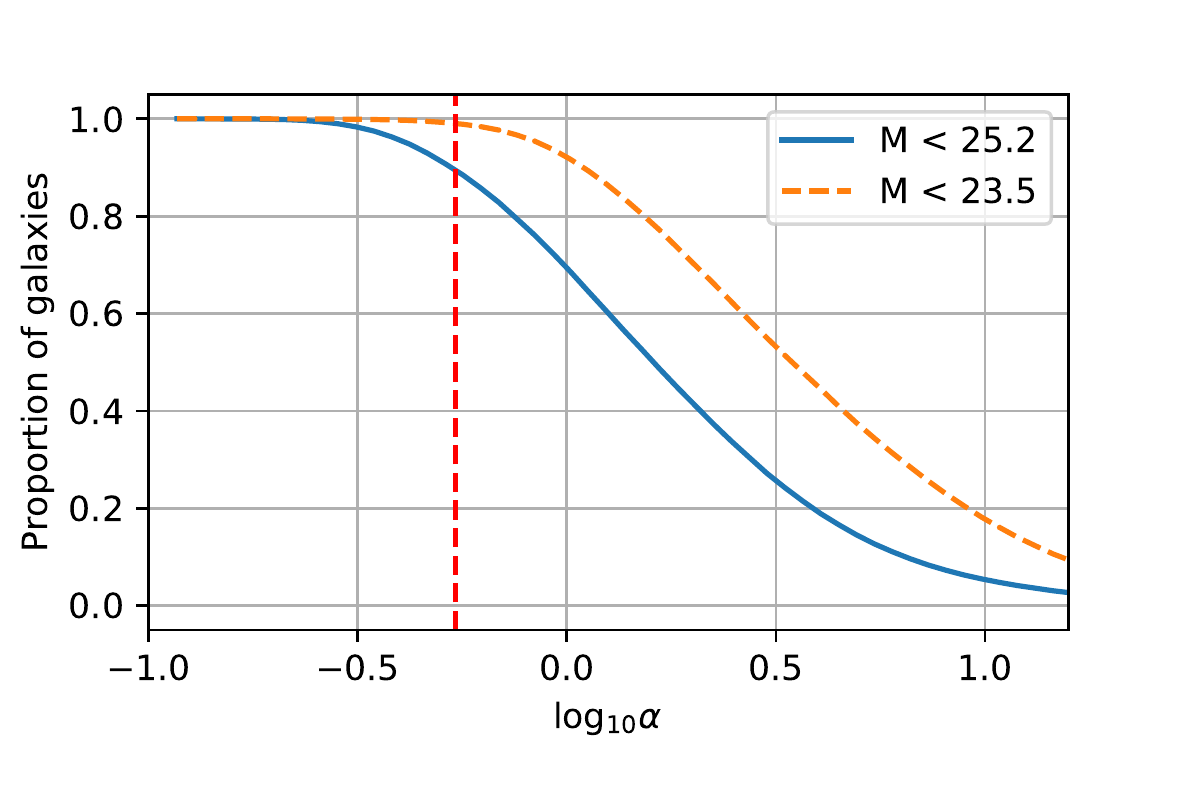}
\end{center}
\vspace*{-5mm}
\caption{Proportion of galaxies in the COSMOS $M_{\textrm{F814W}} < 23.5,\,25.2$ magnitude datasets for which the inner half-light-radius remains unmasked as a function of the image coordinate rescaling factor $\alpha$. The vertical line marks the value $\alpha = 0.542$ chosen for our simulations.}
\label{fig:cosmos_signal_noise_tradeoff}
\end{figure}

Assuming an isotropic Gaussian profile for the galaxy intensities in the target images

\begin{equation}
I(r) = \frac{F}{2\pi \sigma^2}e^{-\frac{r^2}{2\sigma^2}},
\end{equation}
parametrized by a total galaxy flux $F$ and a Gaussian width $\sigma$. Using the Gaussian half-light-radius, $r_{1/2} = \sqrt{2\ln 2} \sigma$, we can express the cutoff radius of a particular galaxy, $r_c$, at which $I(r_c) = I_c$, as follows

\begin{equation}
\label{eq:masked_hlrs}
r_c^2 = \left[\ln\left(\frac{F}{r_{1/2}^2}\right) - \ln I_c + \ln\left(\frac{\ln 2}{\pi}\right) \right] \frac{r_{1/2}^2}{\ln 2}.
\end{equation}
We also assume the following scaling relation between the background RMS in the source and target images

\begin{equation}
\label{eq:sigma_scaling}
\sigma_\mathrm{bg,t} = \frac{\alpha P_t}{P_s
}\sigma_\mathrm{bg,s}
\end{equation}
where $P_s$ and $P_t$ are the pixel-scales of the source and target stamps respectively. This is consistent with the way we scale surface brightness, and assumes the background noise in the source pixels is uncorrelated. Equation \ref{eq:sigma_scaling} however does not take the effects of the target PSF into account. Defining the per pixel cutoff signal-to-noise ratio in the target images, $\mathrm{SN}_c \equiv I_c / \sigma_\mathrm{bg,t}$, then for pixel-scales $P_s = 0.03\arcsec$, $P_t = 0.065\arcsec$, a measured mean background RMS in the source COSMOS stamps of $\sigma_\mathrm{bg,s} \approx 2.9\times 10^{-3}$, and the requirement $\mathrm{SN}_c = 3.5$, we find that the cutoff intensity is $I_c \approx 2.2 \times10^{-2}\,\alpha$. Using equation \ref{eq:masked_hlrs} and the $F - r_{1/2}$ distribution of the COSMOS datasets (scaled to the target stamps), we plot in Figure~\ref{fig:cosmos_signal_noise_tradeoff} the proportion of galaxies for which at least the inner half-light-radius will not be masked (those having $r_\mathrm{c} > r_{1/2}$), as a function of the rescaling factor $\alpha$. Finally, we choose a somewhat arbitrary rescaling factor $\alpha = 0.542$ so that a large fraction of the COSMOS samples ($99\%$ of the $M_{\textrm{F814W}} < 23.5$ dataset and $92\%$ of the $M_{\textrm{F814W}} < 25.2$ dataset) have $r_\mathrm{c} > r_{1/2}$. Moreover, the very low surface brightness galaxies in COSMOS, which are severely masked, are rarely included in the resampled distribution discussed in \S\ref{sec:mag_rad_matching}. 

To summarize this section -- this choice of geometric rescaling factor, between source stamps and target stamps, allowed us to control the noise properties of the simulations (mask pixels having S/N below $3.5$) while minimizing the extent to which stamps are masked. In the following section, the CLASH magnitude-radius distributions are then matched by resampling the simulated distribution. This is done after taking this rescaling factor into account.

\subsection{Matching magnitude-radius distributions}
\label{sec:mag_rad_matching}
\begin{figure*}
\includegraphics[width=\textwidth]{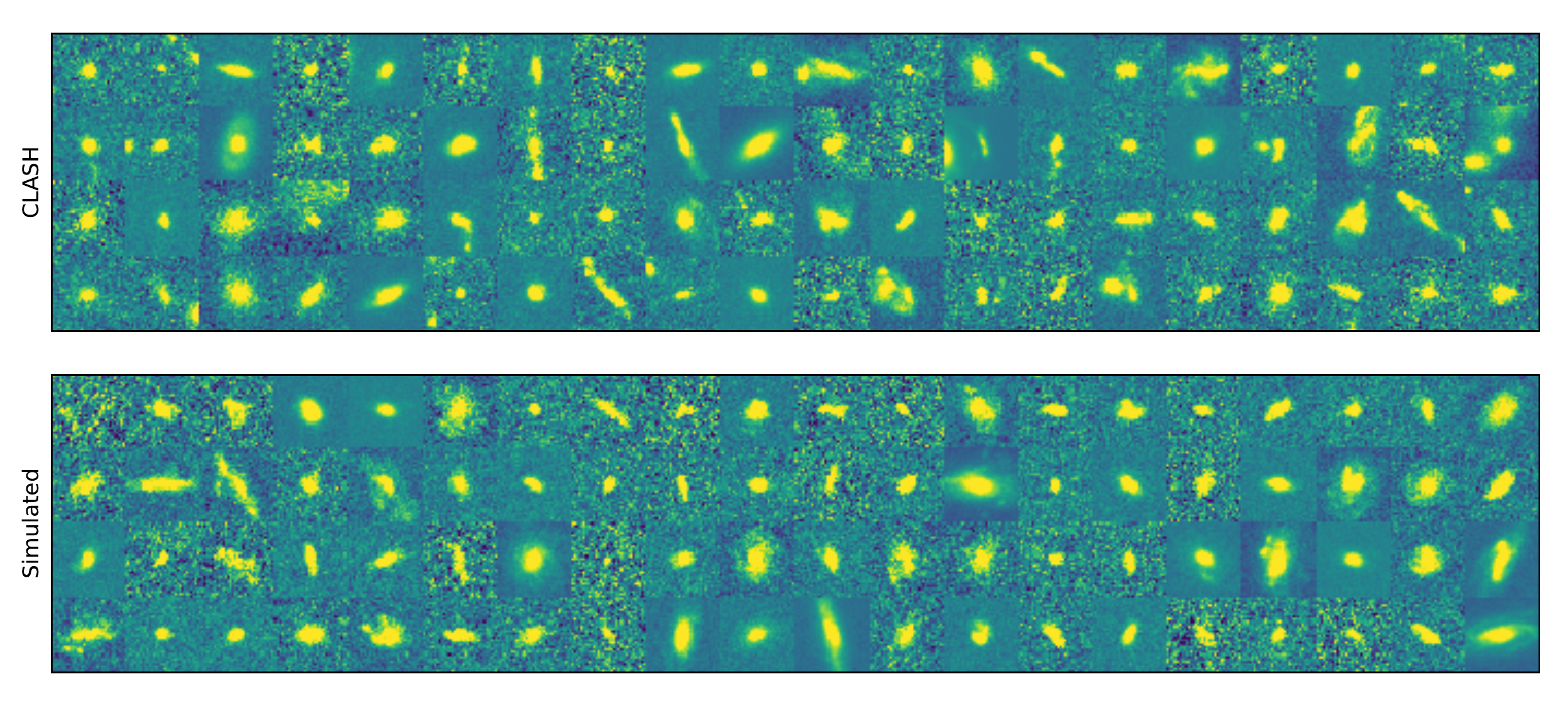} 
\vspace*{-3mm}
\caption{Random subsets of $32 \times 32$ pixel galaxy stamps drawn from the CLASH ACS total image dataset (top) and from the COSMOS based simulations (bottom). Each stamp is shown here after subtracting its original mean and dividing by its standard deviation.}
\label{fig:cosmos_to_clash_flux_stamps}
\end{figure*}

The distribution of observed galaxies is highly diverse and dependent on some well known parameters. Examples are the dependence of morphology on age, the dependence of spectra on redshift, background levels and their effect on S/N conditions, only to name a few. Attempting to simulate the distribution of CLASH \textit{total ACS} filter stamps using the rescaled COSMOS F814W filter observations, and unable to do so perfectly, we choose to focus on matching the COSMOS and CLASH distributions only in the observed magnitude-radius plane. We find that this choice captures much of the variation in appearance (see Figure~\ref{fig:cosmos_to_clash_flux_stamps}). Treating the rescaled COSMOS stamps (see \S\ref{sec:sn_tradeoffs}) as noiseless observations and adding stochastic correlated background noise (as measured on CLASH in \S\ref{sec:correlated_noise}), results in simulations which also reproduce the observed S/N distribution.

\begin{figure}
\begin{center}
\includegraphics[width=\columnwidth]{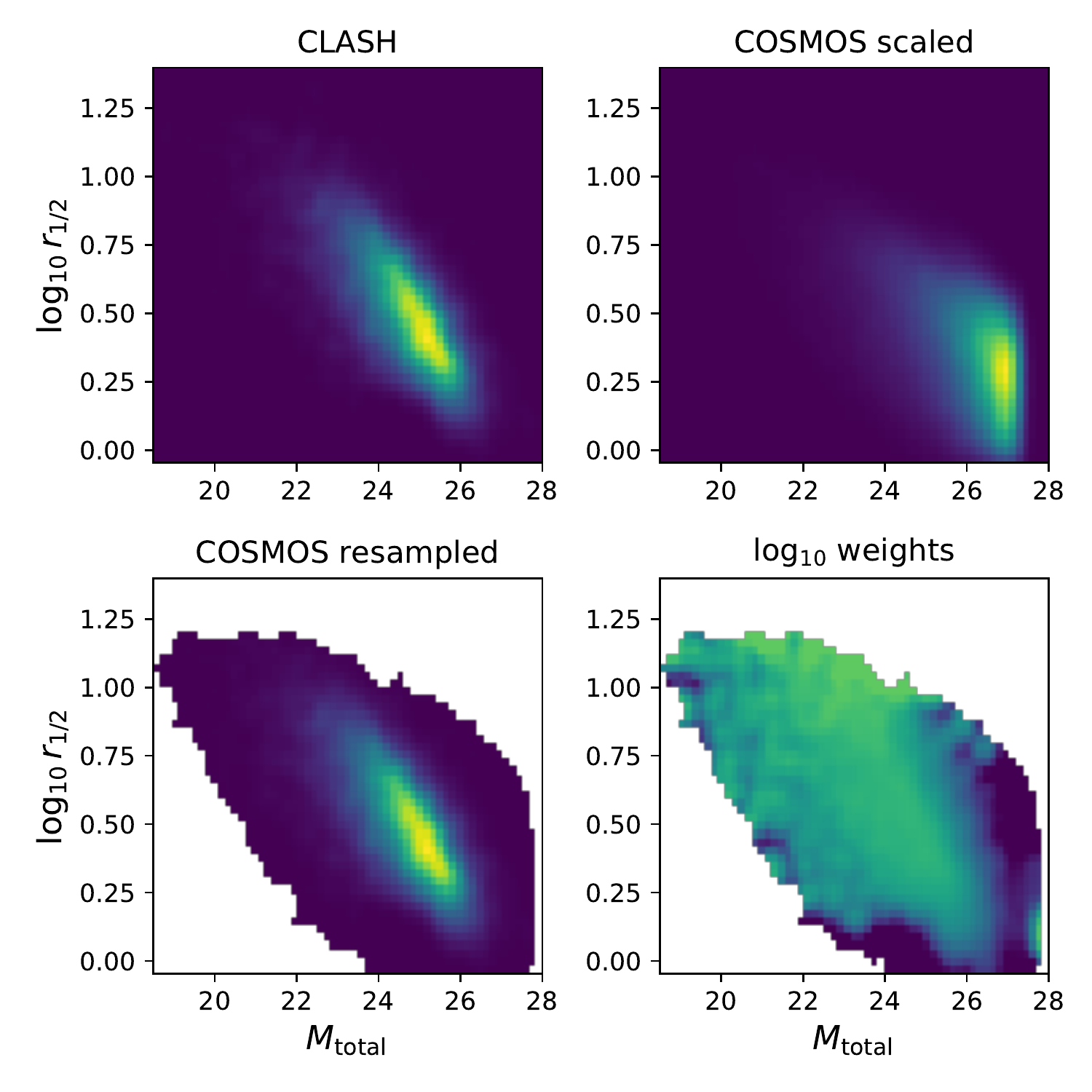}
\end{center}
\vspace*{-3mm}
\caption{Distribution of the CLASH sources (top left), rescaled COSMOS $M_{\textrm{F814W}} < 25.2$ dataset (top right) and this rescaled + resampled COSMOS dataset (bottom left) in the $\log_{10} r_{1/2}\,-\,M_\textrm{total}$ plane. The resampling weights, matching COSMOS to CLASH, are shown on the bottom right.}
\label{fig:cosmos_to_clash_weights_2d}
\end{figure}

\begin{figure}
\includegraphics[width=\columnwidth]{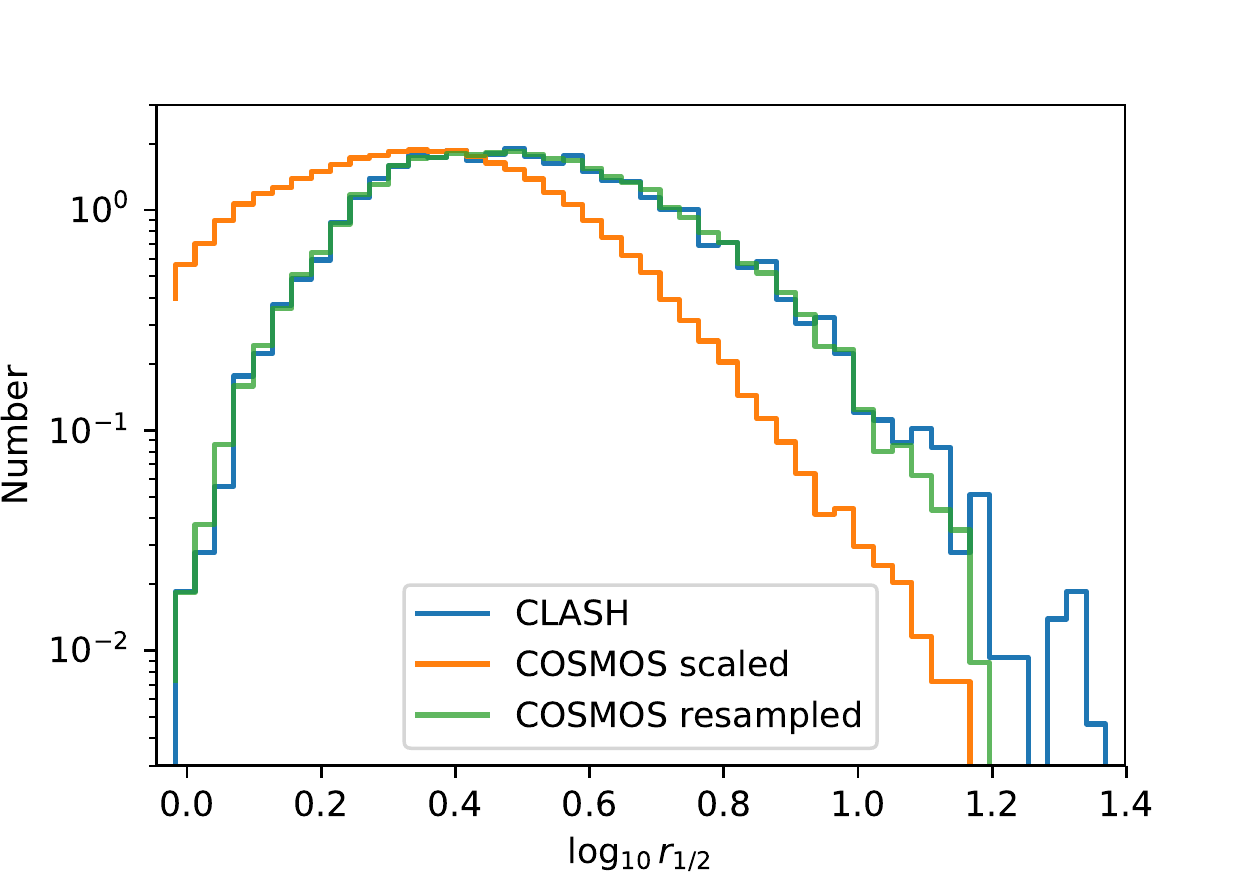}
\vspace*{-3mm}
\caption{Normalized histograms in $\log_{10} r_{1/2}$ bins of the CLASH sources, the rescaled COSMOS $M_{\textrm{F814W}} < 25.2$ dataset and the rescaled + resampled COSMOS $M_{\textrm{F814W}} < 25.2$ dataset.}
\label{fig:cosmos_to_clash_weights_flux_radius}
\end{figure}

\begin{figure}
\includegraphics[width=\columnwidth]{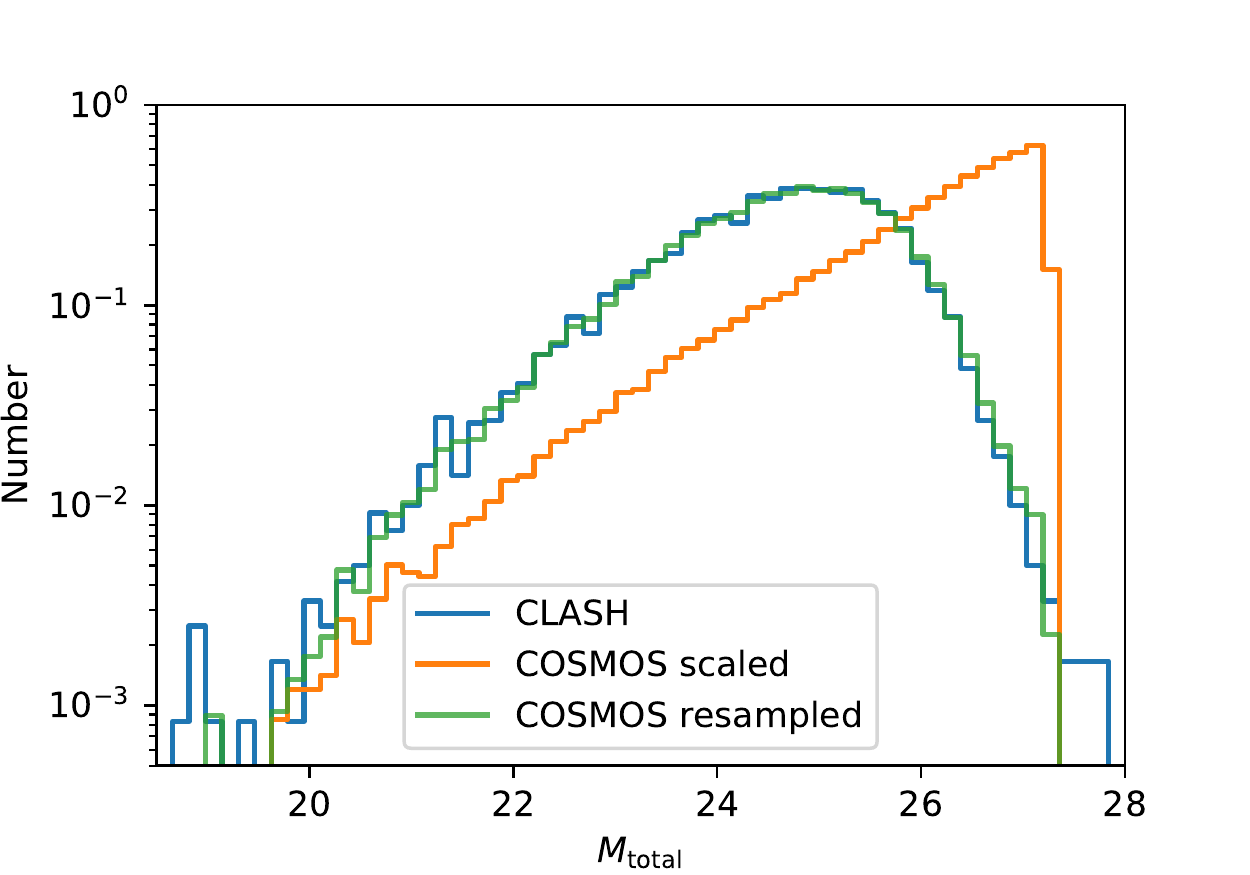}
\caption{Normalized histograms in $M_\mathrm{total}$ bins of the CLASH sources, the rescaled COSMOS $M_{\textrm{F814W}} < 25.2$ dataset and the rescaled + resampled COSMOS $M_{\textrm{F814W}} < 25.2$ dataset.}
\label{fig:cosmos_to_clash_weights_magnitude}
\end{figure}

We perform this matching by resampling the rescaled (as in \S\ref{sec:sn_tradeoffs}) COSMOS stamps. In Figure~\ref{fig:cosmos_to_clash_weights_2d} we show the CLASH observations in the $\log_{10} r_{1/2}\,-\,M_\textrm{total}$ plane (top left), the distribution of the rescaled COSMOS $M_{\textrm{F814W}} < 25.2$ dataset (top right) and the distribution of this resampled COSMOS dataset (bottom left). The rescaling of COSMOS included resizing by the geometric factor $\alpha = 0.542$ and translating the COSMOS distribution in the positive horizontal direction to maximize the overlap with the CLASH distribution (after matching zeropoints and scaling COSMOS flux by $\alpha^2$, this amounted to an extra magnitude shift of $\Delta M = 1.236$). We plot the original CLASH and resampled COSMOS marginal distributions in Figures~\ref{fig:cosmos_to_clash_weights_flux_radius} and \ref{fig:cosmos_to_clash_weights_magnitude}. Aside from a slight lack of simulated galaxies at the high $r_{1/2}$ tail, we see that the distributions match well overall.

\section{Model}
\label{sec:model}
We approach the weak lensing shear estimation problem as a high-dimensional regression task, where given a $32 \times 32$ pixel stamp of the observed galaxy intensities, encoded as a 1024-dimensional vector denoted $x$, we want to predict the two component vector $g$ parametrizing the reduced shear it has undergone. To this end, we wish to find a function $\hat{g} = \Phi(x; w)$ parametrized by a constant weight vector $w$, that would map observations $x$ to consistent and efficient estimates $\hat{g}$ of the true reduced shears $g$, for various pairs $(x,g)$ arising from a joint distribution $(x,g)\sim \mathcal{D}_\mathrm{train}$. We discuss the class of functions $\{\Phi(x; w)\}_w$ which we consider in \S\ref{sec:arch}, and the way in which we search for appropriate weights in \S\ref{sec:train_and_aug}. Finally in \S\ref{sec:test_aug} we discuss how imposing certain symmetry properties on the model further enhances its performance.

The class of functions we use, the optimization algorithm we use to search for a good set of weights, and the way we do this directly from a large set of \textit{training examples}, are all drawn from an approach to machine learning known as \textit{deep learning} \citep[see][]{lecun2015deep}. This approach has in recent years led to considerable progress in the field of computer vision, most notably in visual object classification and detection tasks. Typically, weak lensing shear estimation methods are designed by directly defining a procedure that localizes galaxies (finds their centroid and effective radius) and then measures certain morphological features of the observed galaxies that correlate to the reduced shear they have undergone. The way in which localization is performed, the particular way in which galaxy shape is measured (e.g. RRG's weighted moment approach), the way in which background subtraction is performed, as well as various other observational conditions (e.g. per galaxy S/N), all affect the results of the shear estimation process. The typical approach to these complicating factors is to use matching simulations to calibrate the effects these design choices and observational conditions have on estimation bias \citep[see e.g.][]{fenech2017calibration,samuroff2017dark,zuntz2018dark,mandelbaum2018weak}. In this work the simulations are used not only to calibrate bias, but also to directly learn the estimation algorithm itself using a large set of training examples drawn from the simulations. This is motivated by the success of this type of \textit{end-to-end learning} in finding models that perform better than their manually designed counterparts.\\

The model we describe in this section, as well as its optimization and testing procedures were implemented using the \textsc{TensorFlow}\footnote{\url{https://tensorflow.org}} and \textsc{Keras}\footnote{\url{https://keras.io}} open source deep learning libraries \citep{abadi2016tensorflow,chollet2015keras}. The reader is referred to \S\ref{sec:reproducibility} for further implementation details.

\subsection{Architecture}
\label{sec:arch}
The regression function that we seek, $\Phi(x; w)$, has a particular form known as a multi-layered (or \textit{deep}) \textit{neural network} (DNN), or, more specifically, a deep \textit{convolutional neural network} (CNN). This type of function can be expressed as a composition of $n$ intermediate functions (or \textit{layers}),
\begin{eqnarray}
\Phi(x; w) &=& \Phi_n \circ \Phi_{n-1} \circ ... \circ \Phi_1(x), \\
\Phi_i \circ \Phi_{i-1}({}\cdot{}) &\equiv& \Phi_i(\Phi_{i-1}({}\cdot{}; w_{i-1}); w_i), \\
w &\equiv& \{w_i\}_{i=1}^n.
\end{eqnarray}
Here we used ``$\circ$'' to denote function composition and ``$\,\cdot{}\,$'' to denote unnamed function arguments (e.g., the composition of functions $f(\,\cdot{}\,)$ and $g(\,\cdot{}\,)$ is then $h = f \circ g$ such that $h(x) = f(g(x))$ for some input $x$).

Each intermediate function $\Phi_i({}\cdot{}; w_i)$, parametrized by its own set of weights $w_i$, transforms one intermediate representation $x_i$ of the original input to the following
\begin{equation}
x_{i+1} = \Phi_i(x_i; w_i).
\end{equation}
The i'th \textit{layer} of a DNN refers either to $x_{i}$, or to $\Phi_i({}\cdot{}; w_i)$, depending on the context. The specification of the functions $\Phi_i({}\cdot{}; w_i)$, for yet unspecified values of the weights, constitutes the \textit{architecture} of the DNN. In our CNN, these layer functions are either \textit{convolutional} or \textit{affine}. In Table~\ref{tbl:arch} we list the particular six-layer architecture used in this work. This architecture, computed left-to-right and top-to-bottom, transforms a 1024-dimensional input stamp to a two-dimensional shear estimate $\hat{g}$. The architecture consists of three \textit{convolutional layers} followed by three \textit{affine layers} (refer to Appendix~\ref{sec:layers} for a definition of \textit{convolutional} and \textit{affine layers}, as well as some other operations mentioned in Table~\ref{tbl:arch}). Table~\ref{tbl:arch} defines the dimensions ($n$, $m$) and number ($k$) of convolution kernels used in a convolutional layer by $\texttt{Conv}\,(n,m)\times k$ and the following spatial downsampling operation by $\texttt{Stride}\,(\Delta_x,\Delta_y)$. Affine layers, having output dimension $l$, are denoted by $\texttt{Affine}\,(l)$. Table~\ref{tbl:arch} also lists, for each layer, the dimensions of the intermediate representation $x_i$, as well as the number of scalar parameters included in the weight vector $w_i$. In total, our model's architecture includes $107\,866$ parameters which are subsequently optimized (or trained), using the simulation training set, in \S\ref{sec:train_and_aug}. We remark here that at the present time, there is no rigorous way of finding an optimal architecture for a particular deep learning problem. Our architecture was chosen after testing only a few alternatives. It is therefore likely that other architectures, leading to better performance, can be found.

\input{arch.tex}

\subsection{Training and data augmentation}
\label{sec:train_and_aug}

For the purpose of optimizing the model, we use the standard mean-squared-error (MSE) \textit{loss function}, evaluated on a distribution of (stamp, shear) pairs

\begin{equation}
\label{eq:mse_loss}
L(w) = \E_{(x,g)\sim\mathcal{D}_\mathrm{train}}\left[\lvert\lvert\Phi(x; w) - g \rvert\rvert^2\right],
\end{equation}
where $\mathcal{D}_\mathrm{train}$ is the distribution we draw our training samples from. The global minimum $w^*$ of such a loss function, for an expressive enough model, should result in an estimator $\Phi(x; w^*) = \E\left[g|x\right]$, which produces the conditional mean of the shear $g$ given the observation $x$ \citep[see e.g.][]{bishop1994mixture}. In practice, the optimization is performed using a finite set of training examples, using the specific architecture defined in \S\ref{sec:arch} and using an iterative stochastic optimization algorithm known as \textsc{RMSProp} \citep{tieleman2012lecture}, which is not guaranteed to produce a global minimum. The training set we use at this stage differs from the simulations as defined in \S\ref{sec:simulations} in the following ways:
\begin{enumerate}
\item Both the resampled and the original COSMOS datasets discussed in \S\ref{sec:simulations} have a large over-abundance of high magnitude galaxies. We found that training using a more homogeneous magnitude distribution enhances the model's performance on the CLASH matched simulations. We homogenize the training set by sampling an equal number of galaxies in each of the original COSMOS $M_{\mathrm{F814W}}$ magnitude segments: $[16, 19.5]$, $[19.5, 21.5]$, $[21.5, 23.5]$.
\item The scale of simulated stamps is randomly and uniformly augmented in the range $\pm 15\%$.
\item The post shear centroid position of simulated stamps is randomly and uniformly augmented in the range $\pm 7\,\textrm{px}$ both in the $x$ and $y$ directions.
\item Stamps are randomly mirrored (with probability 0.5), and rotated randomly and uniformly by an angle $\theta \in \left[0,2\pi\right]$.
\end{enumerate}
The augmentations described here allow us to enlarge the effective size of the training and test sets and also encourage the resulting model to be invariant to the above transformations, specifically to translation (see \S\ref{sec:sensitivity}). Additionally, without enlarging the training data using this data augmentation procedure the resulting models tend to over-fit the training dataset and perform poorly on the test dataset. This property, known as \textit{over-fitting} (or lack of generalization), commonly occurs in machine learning models having a large number of parameters and a relatively small number of training examples. This is why these augmentations are necessary. Note that although the training set includes augmentations that may be leading to an "improper" population resampling, the test set, with which we calibrate and evaluate the performance of our model in \S\ref{sec:performance_sim}, does not include these type of augmentations (i - iii) and is properly sampling the population as defined in \S\ref{sec:simulations} in these respects.\\

Table \ref{tbl:sim_datasets} gives a quantitative summary of how the original COSMOS $M_{\textrm{F814W}} < 23.5,\,25.2$ magnitude datasets were split and enlarged into the final training and test sets.  Both the training stamps and the test stamps are normalized, before the CNN operates on them, by subtracting the stamp's per-pixel mean and dividing by its standard deviation (see e.g. Figure~\ref{fig:cosmos_to_clash_flux_stamps}). Doing this both assists in the convergence of the optimization and has the added benefit of making the trained model invariant to the subtracted background level. Correctly estimating the background level is a difficult task in itself and an incorrect estimate can in turn affect RRG moment estimates (see discussion in \S\ref{sec:sensitivity}). Having the shear estimator insensitive to constant background levels is therefore an advantageous property.

The CNN estimator is also completely invariant to flat back- ground levels subtracted from the stamp. As discussed in §4.2, this is due to the way stamps are normalized before the CNN operates on them. Accurately measuring background levels in galaxy clus- ter lensing is particularly challenging due to the spatially varying foreground intra-cluster-light (ICL). Attempts have recently been made to better model the ICL and correct for its systematic effect on background source magnitude and photometric redshift measure- ments (see Molino et al. 2017; Gruen et al. 2018). We expect an incorrect background estimation to also affect RRG shear estimates through its affect on the SExtractor segmentation process (and half-light-radius estimation), as well as through the weighted mo- ments evaluated by RRG. This insensitivity of the CNN estimator to constant background levels is therefore an advantageous property.

At the final stage of training, we set all model weights constant and fit the last affine layer by linear-least-squares regression between the $30$-dimensional representation at layer five and the $2$-dimensional output at layer six. This is done using the training set in which only augmentation (iv) was performed. This allows us to tune the model to the CLASH matched distribution. Additionally, we multiply all weights in layer six by $\left[1-L(w^*)/\textrm{Var}(g)\right]^{-1}$, where $w^*$ are the optimized weights and $L(w^*)$ is the MSE loss evaluated on the CLASH matched training set at $w^*$. This is done to correct for the effects of \textit{regression attenuation} as discussed in Appendix~\ref{sec:regression_attenuation}.

\input{sim_datasets.tex}

\subsection{Enforcing shear symmetries}
\label{sec:test_aug}

The components of the true reduced shear transform as $[g_1,g_2]^T \rightarrow R(2\theta)[g_1,g_2]^T$ in response to a rotation of the image coordinates by an angle $\theta$, and as $[g_1,g_2] \rightarrow [g_1,-g_2]$ under reflection. Although we train the model in \S\ref{sec:train_and_aug} using a distribution of samples containing both mirrored and rotated copies of the original COSMOS datasets, the optimization does not necessarily converge to a model estimator having these precise transformation properties. We impose these transformation properties on our trained estimator in the following way
\begin{equation}
\tilde{g}_{\textsc{cnn}} \equiv \phi(x; w^*) = \frac{1}{8}\sum_{i\in\{-1,1\}}\sum_{j=0}^3 t_{i,j}(\Phi(T_{i,j}(x); w^*)).
\end{equation}
Here the function $T_{i,j}(x)$ vertically flips the stamp $x$, depending on the value of $i$, and rotates it by an angle of $\pi j / 2$. The function $t_{i,j}(g)$ performs the appropriate shear transformation (as discussed above). Using $\phi(x; w^*)$ instead of $\Phi(x; w^*)$ as our estimator results in an additional overall reduction of ${\sim}2.5\%$ in the RMS errors measured in \S\ref{sec:perf_overall}.

\section{Performance on simulated data}
\label{sec:performance_sim}

Based on the simulations described in \S\ref{sec:mag_rad_matching}, we measure the performance of an estimator $\hat{g}_{e}$ using the RMS error $\sigma_{e,i}$,

\begin{eqnarray}
\sigma_{e,i}^2 &=& \E_{(x,g)\sim\mathcal{D}_{\mathrm{test}}}\left[\left(\hat{g}_{e,i} - g_i \right)^2\right], \\
\hat{g}_{e,i} &=& (\tilde{g}_{e,i} - c_{e,i})/(1+m_{e,i}), \label{eq:bias_correction}
\end{eqnarray}
where we denoted the true simulated shears by $g$, the uncalibrated estimates by $\tilde{g}_{e,i}$, the calibrated estimates by $\hat{g}_{e,i}$, the type of estimator used by subscript $e \in \{\textsc{cnn},\textsc{rrg}\}$, the reduced shear component by subscript $i \in \{1,2\}$, and the specific test distribution we evaluate performance on using $\mathcal{D}_{\mathrm{test}}$. Here we use our own implementation of the RRG algorithm as it is described in \cite{rhodes2000weak}. The PSF we use in the simulations, for all generated galaxies is an isotropic Gaussian profile having a FWHM of $0.121\arcsec$ as discussed in \ref{sec:point_sources}. We evaluate the performance on $\mathcal{D}_\mathrm{test}$ only after calibrating an affine bias model for each estimator and component of shear separately,
\begin{multline}
(m_{e,i}, c_{e,i}) = \\
\argmin_{(m_{e,i}, c_{e,i})} \E_{(x,g)\sim\mathcal{D}_{\mathrm{test}}}\left[\left(\tilde{g}_{e,i} - c_{e,i} - (1+m_{e,i})g_i \right)^2\right] .
\end{multline}
We do not use a separate calibration set to do this, but instead use the same test set both to calibrate biases and then to measure RMS errors. By repeated simulation of this calibration process we find for the RMS error estimates (see Table~\ref{tbl:perf_overall}) statistical uncertainties of $\pm1.0\times 10^{-3}$ and $\pm8.6\times 10^{-4}$ for the RRG and CNN estimators respectively and systematic uncertainties below $+5\times 10^{-5}$.\\

By varying the test set $\mathcal{D}_\mathrm{test}$, we measure how each estimator's performance varies with varying source population magnitude (\S\ref{sec:perf_mag}) and assess the sensitivity of the estimators to centroid misalignment and to an improper background subtraction (\S\ref{sec:sensitivity}). In \S\ref{sec:perf_overall} we discuss the overall RMS errors and biases for the CNN estimator when evaluated on the full population of the CLASH matched simulations.

\subsection{Magnitude dependence}
\label{sec:perf_mag}

\begin{figure}
\begin{center}
\includegraphics[width=1.1\columnwidth]{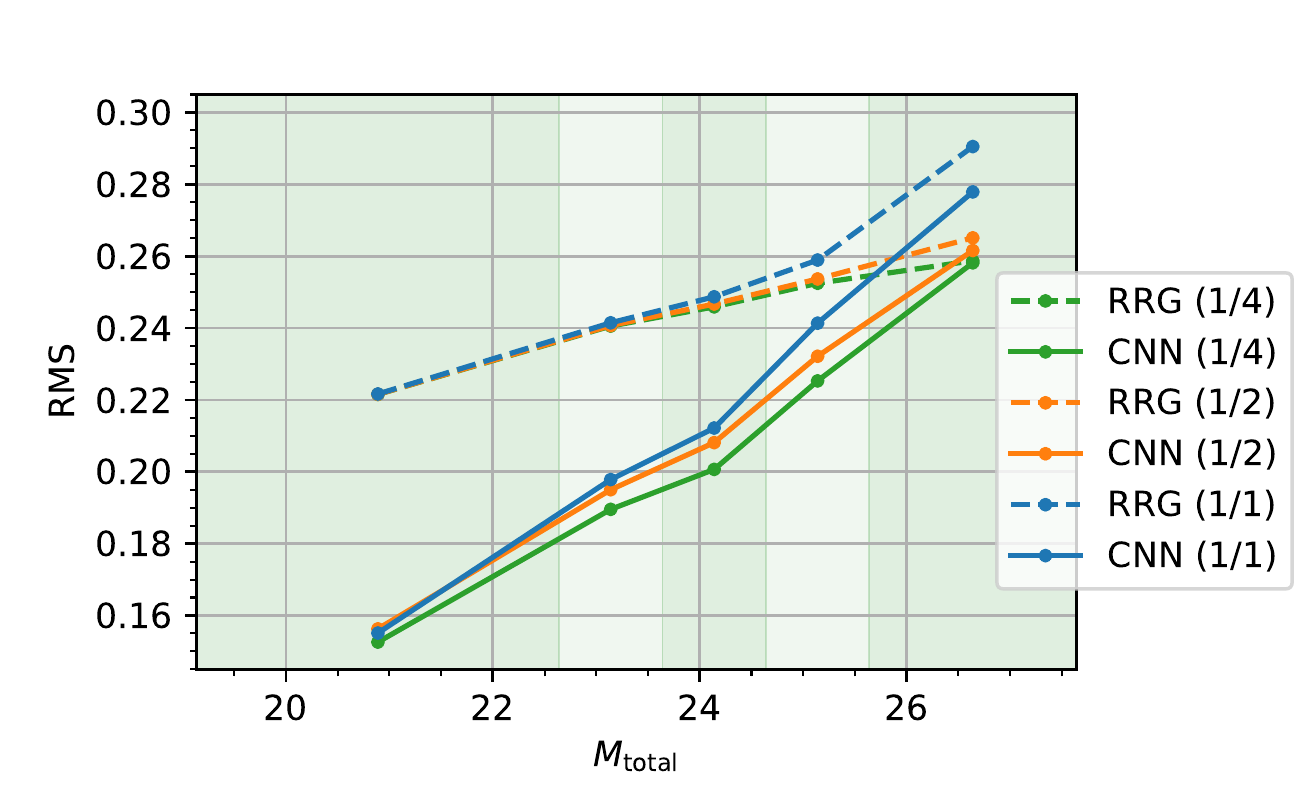}
\end{center}
\caption{Reduced shear RMS error (average of the two shear components) of the CNN and RRG estimators for galaxies simulated in five magnitude bins (vertical strips). We show results using the original CLASH background noise as well as noise reduced by factors of $1/2$ and $1/4$, training a separate CNN estimator in each case.}
\label{fig:performance_per_mag_lower_noise}
\end{figure}

In Figure~\ref{fig:performance_per_mag_lower_noise} we measure performance in five $M_{\mathrm{total}}$ magnitude bins. We do this with background noise simulated using the original noise reproducing kernel $K_{\textsc{bg}}$ (see \S\ref{sec:correlated_noise}), and using kernels $K_{\textsc{bg}}/2$ and $K_{\textsc{bg}}/4$ to simulate lower noise conditions. This generates stamps with additive background noise having the original CLASH standard deviation, and standard deviations reduced by factors of $1/2$ and $1/4$. Different CNN models were trained for each noise level. We see that the RMS errors of both the RRG and CNN estimators rise with magnitude and that higher background noise levels lead to higher RMS errors, mainly at the fainter end. The RRG error curve also appears to converge, at low noise levels, to a bottom envelope. 

In the brighter galaxies, more substructure is typically visible (see Figure~\ref{fig:cosmos_to_clash_flux_stamps}). This substructure is what presumably provides the CNN estimator with additional shear information. The RRG estimator only has access to the second order surface brightness moments and is therefore limited in its statistical efficiency by their distribution.

\subsection{Sensitivity to alignment and background subtraction}
\label{sec:sensitivity}
\begin{figure}
\includegraphics[width=\columnwidth]{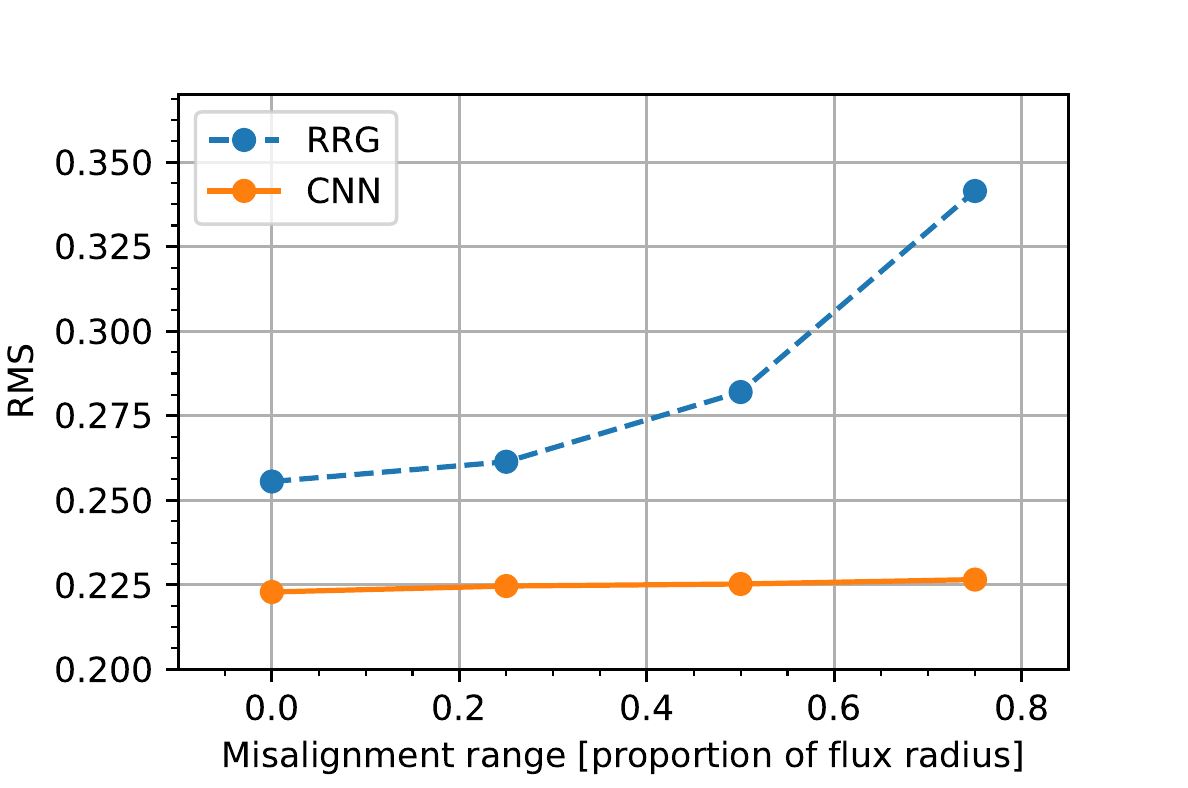}
\caption{Reduced shear RMS error (average of the two shear components) of the CNN and RRG estimators for galaxies simulated with an increasingly perturbed centroid.}
\label{fig:sensitivity_to_translation}
\end{figure}

In \S\ref{sec:perf_mag} we did not include the effects that centroid determination has on shear RMS errors. The centroid used there was the one provided by the COSMOS sample. This sample has higher resolution and higher S/N than the CLASH sample we eventually simulate and should therefore have a more accurate centroid measurement (visible in Figure~\ref{fig:cosmos_to_clash_flux_stamps}). Here we would like to separately measure the effects a random misalignment of the galaxy centroid has on each estimator. We do this by producing test samples in which the centroid of each galaxy is randomly perturbed uniformly in a range having a total extent which is a particular proportion of the galaxy's half-light-radius, both in the $x$ and $y$ directions. This is done post shear. The results are presented in Figure~\ref{fig:sensitivity_to_translation}. We see that the RMS errors of the RRG estimator rise significantly above a misalignment (range) of ${\sim}0.5$. The performance of the CNN estimator, on the other hand, is barely affected by perturbations of the centroid in this range. This insensitivity of the CNN estimator is a result of the translational augmentations performed when producing the training set (see \S\ref{sec:train_and_aug}).\\

The CNN estimator is also completely invariant to flat background levels subtracted from the stamp. As discussed in \S\ref{sec:train_and_aug}, this is due to the way stamps are normalized before the CNN operates on them. Accurately measuring background levels in galaxy cluster lensing is particularly challenging due to the spatially varying foreground intra-cluster-light (ICL). Attempts have recently been made to better model the ICL and correct for its systematic effect on background source magnitude and photometric redshift measurements \citep[see][]{molino2017clash,gruen2018dark}. We expect an incorrect background estimation to also affect RRG shear estimates through its affect on the \textsc{SExtractor} segmentation process (and half-light-radius estimation), as well as through the weighted moments evaluated by RRG. This insensitivity of the CNN estimator to constant background levels is therefore an advantageous property.

\subsection{Overall performance and bias}
\label{sec:perf_overall}

\begin{figure}
\begin{center}
\includegraphics[width=1.05\columnwidth]{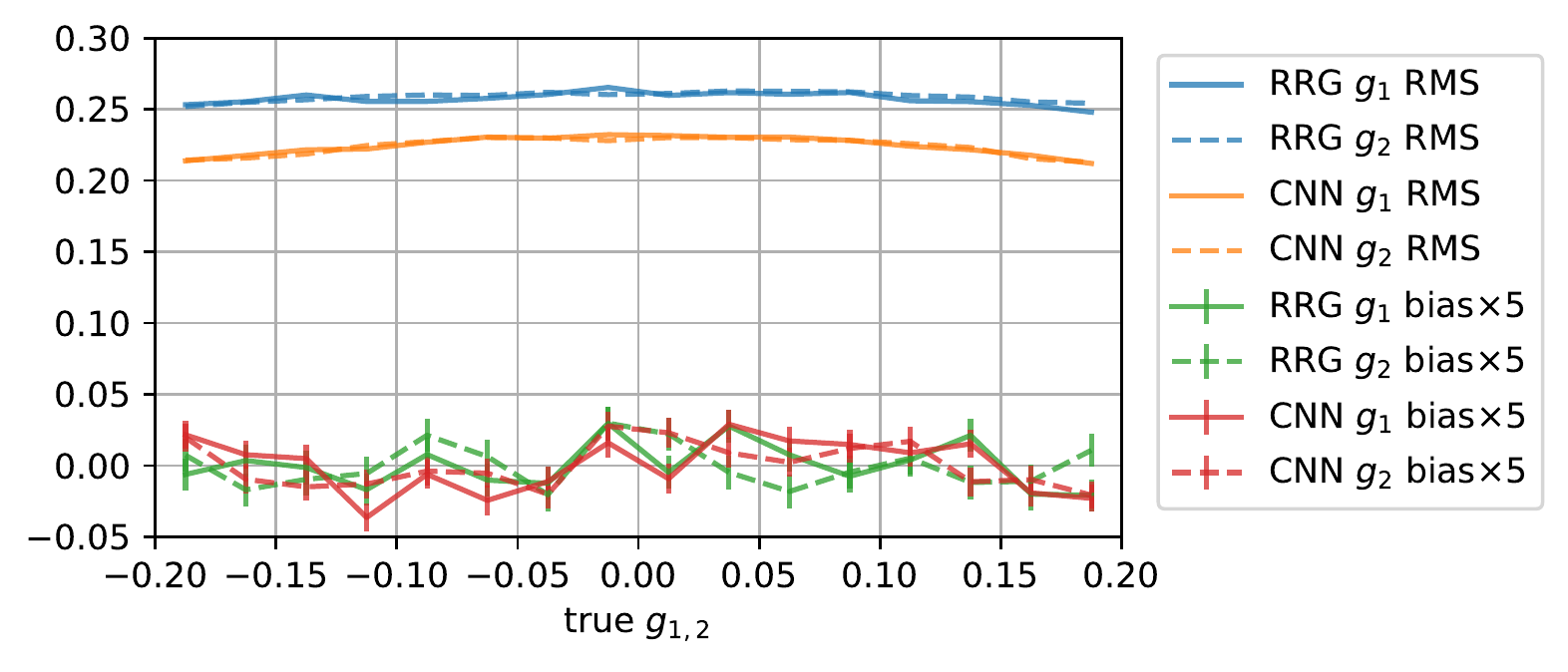}
\end{center}
\caption{RMS errors and biases in 16 separate $g_1$ and $g_2$ true shear bins for the bias corrected RRG and CNN estimators. Each bin contains approximately 12,500 samples from the full CLASH matched test set. The bias correction here is constant and is the one presented in Table~\ref{tbl:perf_overall}. Note that bias here is presented at 5 times the amplitude for clarity.}
\label{fig:performance_per_shear}
\end{figure}

In Table~\ref{tbl:perf_overall} we summarize the calibration parameters and performance of the RRG and CNN  estimators when evaluated on the full CLASH matched test set. As discussed in \ref{sec:sensitivity}, these simulations do not account for a more realistic stamp misalignment or an incorrect background estimation. The simulated shear components in this test set were independently drawn such that $g_i \sim U(-0.2,0.2)$. Overall the RRG estimator has an RMS error approximately 15\% higher compared to that of the CNN estimator, at all tested true shear bins. In Figure \ref{fig:performance_per_shear} we measure the RMS errors and biases in 16 separate $g_1$ and $g_2$ true shear bins, each containing approximately 12,500 samples. We see that the absolute gap in RMS errors between the two methods is approximately constant as a function of shear, with both methods showing higher RMS errors at lower absolute shears. The bias (presented at five times the amplitude for clarity) appears to be consistent with being zero at this confidence level for each of the tested true shear bins.

\input{perf_overall.tex}

\section{CLASH Weak lensing analysis}
\label{sec:clash_analysis}

To measure the statistical performance of the estimators on the CLASH observations we make use of the angular correlations in the reduced shear fields in each cluster. We first compute in \S\ref{sec:smooth_fields} fields smoothed at various length scales. We then measure the covariance properties of the smooth fields for the two estimators. This enables us in \S\ref{sec:relative_bias} to calibrate a relative affine bias model between the CNN and RRG estimates on this dataset. In \S\ref{sec:statistical_errors_clash} we correct for this bias and use a held out set of background galaxies to measure the scatter between the per galaxy point estimates and the smooth fields. This provides us with an evaluation of the per galaxy statistical errors of the RRG and CNN estimators on the CLASH observations. Our final product for the 20 CLASH clusters is then a shear catalogue having RMS errors reduced by approximately 26\%, as well as smooth shear fields having enhanced angular resolutions (at a set noise level).

\subsection{Smooth field regression}
\label{sec:smooth_fields}

\begin{figure*}
\begin{center}
\includegraphics[width=0.95\textwidth]{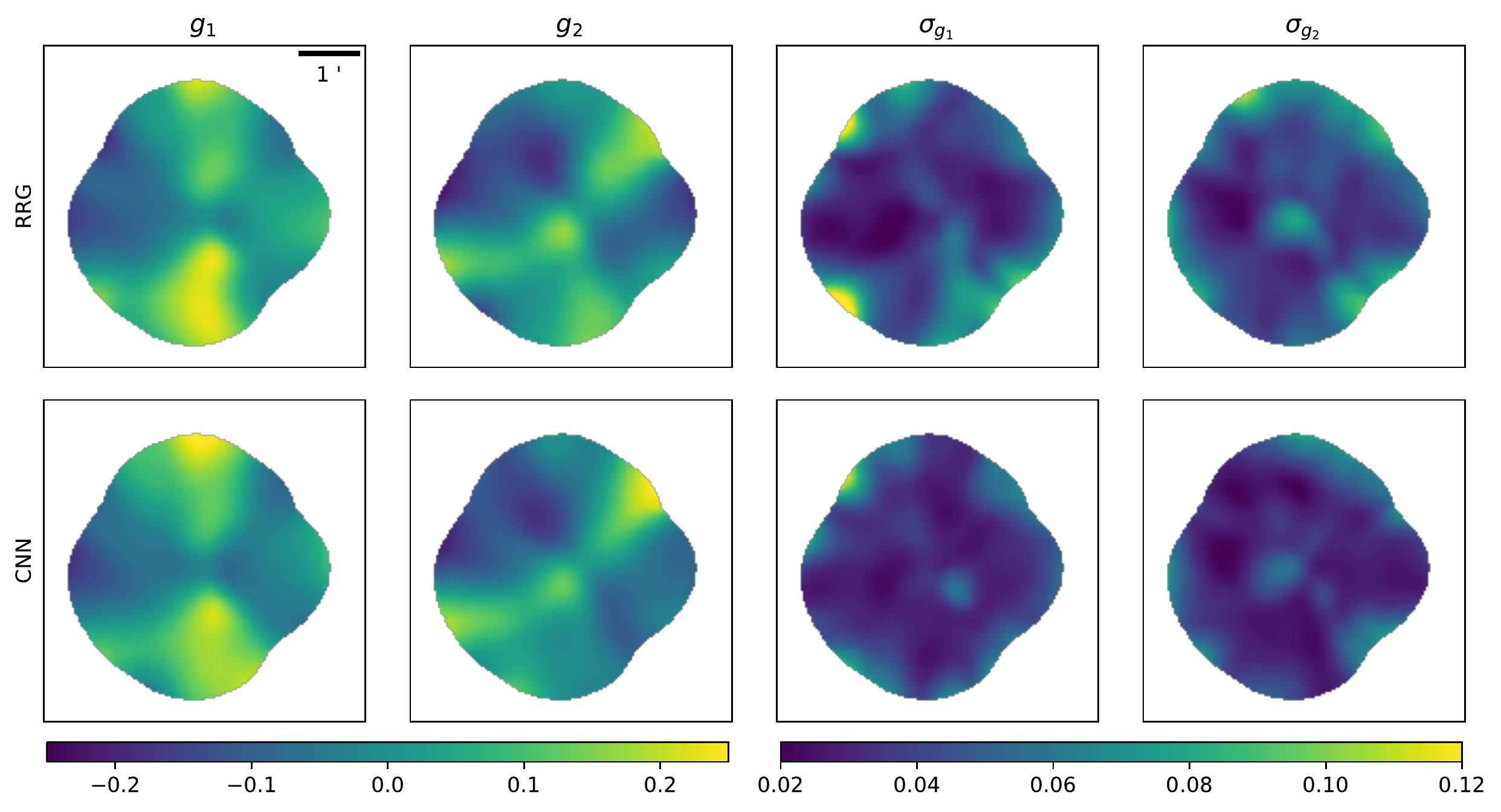}
\end{center}
\caption{Reduced shear field components $g_{i}$, and standard deviations $\sigma_{g_{i}}$, for the RRG and CNN estimators computed using a smoothing length of $\lambda = 19.5\arcsec$ in representative CLASH cluster MACSJ1720+35.}
\label{fig:clash_smooth_fields}
\end{figure*}

For a given galaxy cluster, we measure the reduced shear field $g_{i,e,\lambda}(\vec{\theta})$, smoothed at angular scale $\lambda$ 
\begin{eqnarray}
g_{i,e,\lambda}(\vec{\theta}) &=& \sum_{j=1} \hat{g}^{(j)}_{i,e} \delta(\vec{\theta}-\vec{\theta}_j) * S_\lambda (\vec{\theta}), \\
S_\lambda (\vec{\theta}) &\equiv& \frac{1}{\sqrt{2\pi}\lambda}\exp\left(-\frac{\lvert\lvert \vec{\theta} \rvert\rvert^2}{2 \lambda^2}\right),
\end{eqnarray}
where $\vec{\theta}$ denotes sky coordinates, $\hat{g}^{(j)}_{i,e}$ is the reduced shear estimate of component $i$ computed with estimator $e$ of background galaxy $j$ located at $\vec{\theta}_j$ and $S_\lambda (\vec{\theta})$ is a Gaussian smoothing kernel having width $\lambda$. We define spatial convolution of two images $f(\vec{\theta})$ and $h(\vec{\theta})$ as 
\begin{equation}
(f*h)(\vec{\theta}) = \int \mathrm{d}^2\theta' f(\vec{\theta'}) h(\vec{\theta} - \vec{\theta}').
\end{equation}

\begin{figure*}
\begin{center}
\includegraphics[width=\textwidth]{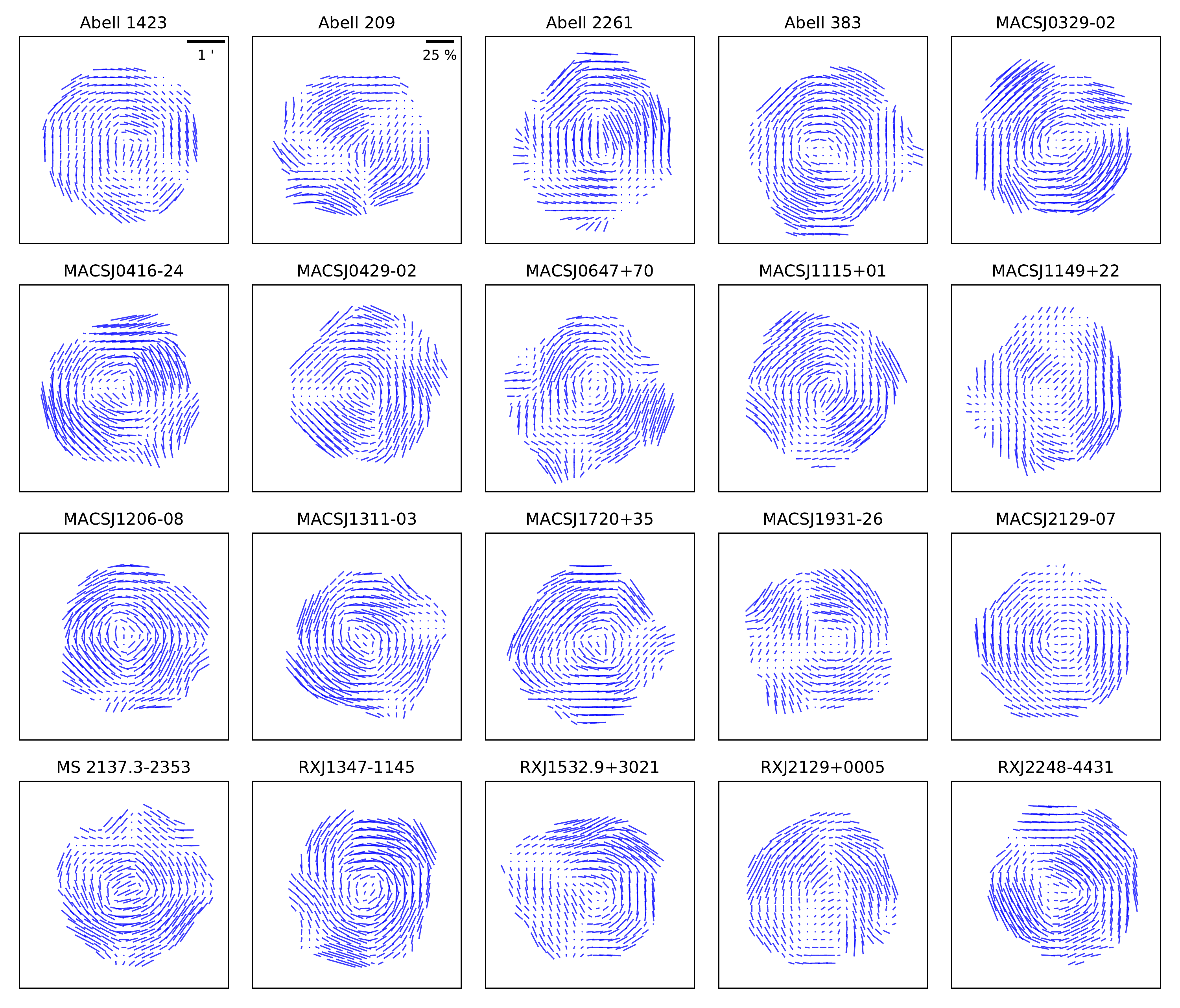}
\end{center}
\caption{Reduced shear fields, $g_{i,\lambda}(\vec{\theta})$, for the CNN estimator, computed using a smoothing length of $\lambda = 19.5\arcsec$ in the 20 CLASH clusters analysed. In all clusters J2000 north is up and both the angular scales and the shear scales are identical. An angular scale-bar corresponding to $1'$ is shown next to Abell 1423 and a scale-bar corresponding to a reduced shear of 25\% is shown next to Abell 209. The lengths and directions of the sticks, at each point, correspond to the magnitude and angle $\phi$ of the complex representation $g_{1,\lambda} + i g_{2,\lambda} = |g_{\lambda}| e^{2i\phi}$ of the reduced shear field.}
\label{fig:clash_wl_rrg_cnn_quiver_mag}
\end{figure*}

In Figure \ref{fig:clash_smooth_fields} we show the resulting reduced shear field component maps $g_{i}$, for the RRG and CNN estimators at a smoothing length of $\lambda = 19.5\arcsec$ for cluster MACSJ1720+35. We also present the field standard deviations $\sigma_{g_{i}}$ estimated using 100 bootstrap resampling iterations of the full background galaxy catalogue. The CNN smooth shear fields for the full set of CLASH clusters are presented in Figure \ref{fig:clash_wl_rrg_cnn_quiver_mag}.\\

\subsection{Relative bias calibration}
\label{sec:relative_bias}
\begin{figure}
\begin{center}
\includegraphics[width=1.0\columnwidth]{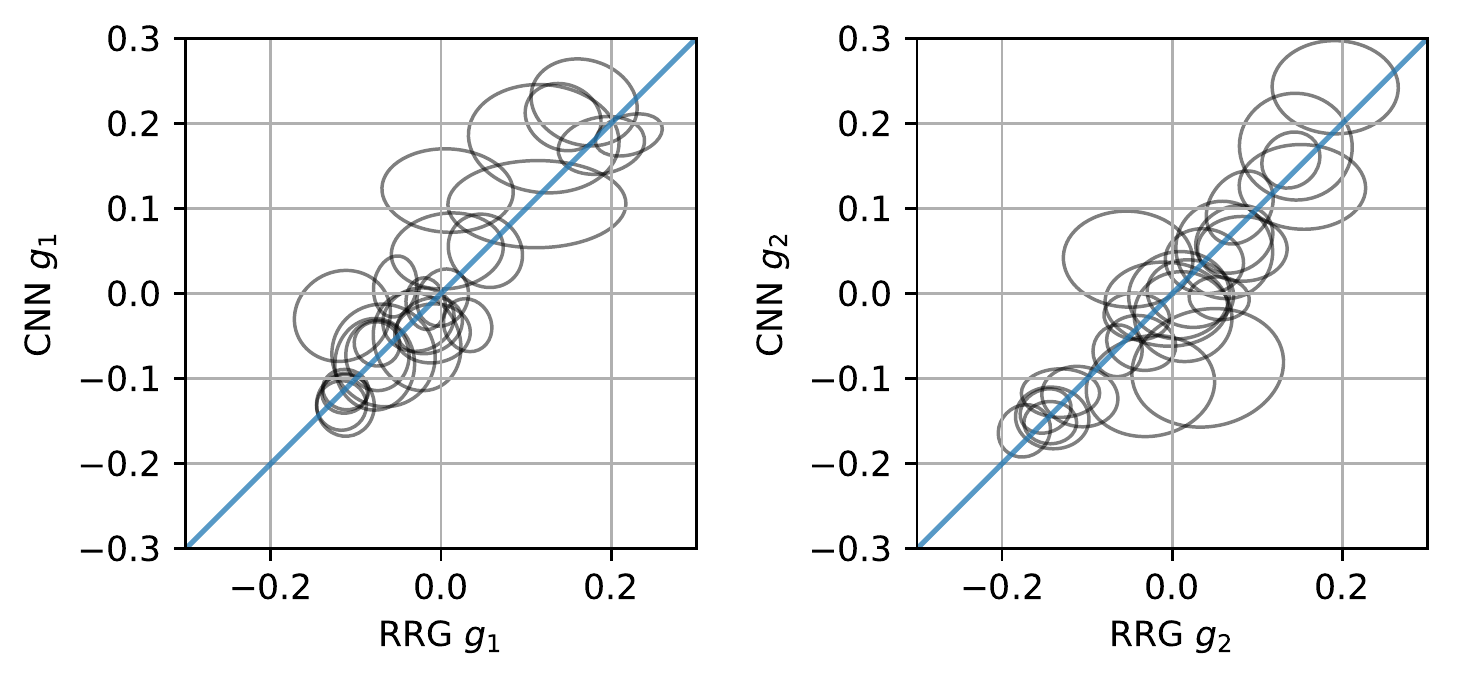}
\end{center}
\caption{CNN vs. RRG shear estimates sampled at a random set of locations (at 31 points) in the CLASH cluster MACSJ1720+35 field. At each point and for each estimator and shear component, smooth field estimates as well as the $1 \sigma$ bootstrap confidence ellipse is shown (a smoothing length of $\lambda = 19.5\arcsec$ was used here).}
\label{fig:clash_biases_specific}
\end{figure}

\begin{figure}
\begin{center}
\includegraphics[width=1.0\columnwidth]{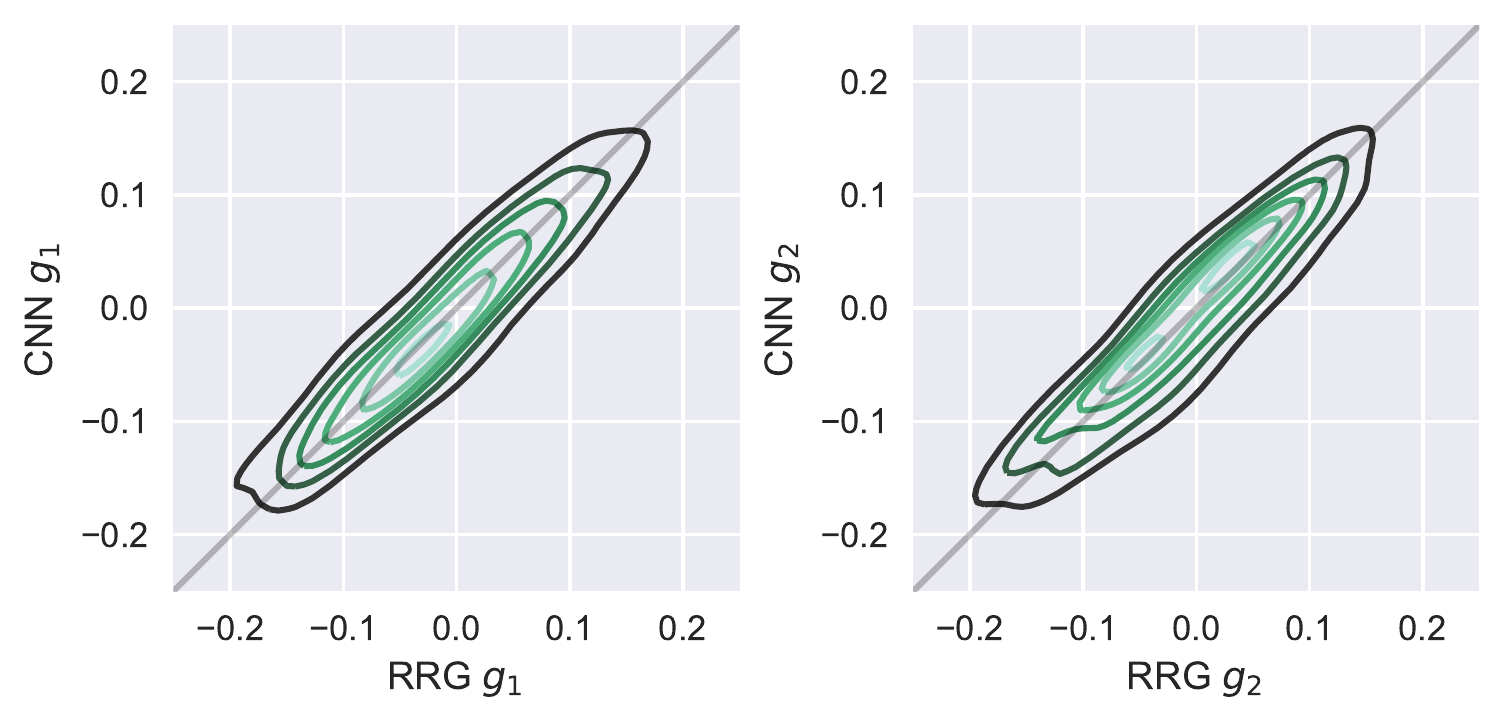}
\end{center}
\caption{Density plots of RRG and (mutually calibrated) CNN smooth ($\lambda = 19.5\arcsec$) shear field sparse samples, for the combined set of CLASH galaxy clusters.}
\label{fig:clash_biases}
\end{figure}

To allow a comparison of the relative performance of the RRG and CNN estimators on real observations we first wish to calibrate an additional relative affine bias model on the CLASH observations. To do so we utilize the per-cluster shear fields $g_{i,e,\lambda}(\vec{\theta})$ and bootstrap error estimates mentioned in the previous section. In these aforementioned bootstrap iterations, we estimate the full $2\times 2$ covariance matrix, $\Sigma_{i,\lambda}(\vec{\theta})$, at each point in the field. For each component of shear, these matrices measure both the variances of each estimator at angular position $\theta$, as well as their mutual covariances there. We randomly sample the shear fields uniformly at $1/10$'th the density of each cluster's available background sources. We do this to minimize the effect noise correlations between the sampled points would have on (underestimating) the bias parameter confidence intervals. Figure \ref{fig:clash_biases_specific} illustrates this process by showing the sparse random sample evaluated in cluster MACSJ1720+35. Using the sparse samples from the full set of CLASH clusters we fit the relative bias model using an error-in-variables likelihood model \citep[\S 7]{hogg2010data}. The resulting relative bias parameters are presented in Table \ref{tbl:clash_biases}. Figure \ref{fig:clash_biases} shows density plots of the RRG and CNN estimates for the full set (over all galaxy clusters) of sparse samples, after applying this additional calibration to the CNN shears.\\

\input{clash_biases.tex}

\subsection{Statistical error analysis}

We wish to measure the statistical errors of each of the (mutually calibrated) estimators in the CLASH fields. To do so we first evaluate the following scatter between the smooth shear field and the shear measured at a particular background source
\begin{equation}
\sigma^2_{\textrm{field-source},i,e}(\lambda) = \E_k\left\{ \E_{j\in S_k}\left[\left(g^{\overline{S}_k}_{i,e,\lambda}(\vec{\theta}_j) - \hat{g}^{(j)}_{i,e}\right)^2\right]\right\}.
\end{equation}
Here the index $k$ denotes the CLASH cluster, $S_k$ is a randomly selected $50\%$ hold-out set of background sources in this cluster and $g^{\overline{S}_k}_{i,e,\lambda}(\vec{\theta}_j)$ are the smooth shear fields evaluated on $S_k$'s complement. Due to this random split of the background sources, we can assume that the smooth field noise and the per source noise are statistically independent. Using 100 bootstrap resampling iterations on the held-in background sources $\overline{S}_k$, we estimate the following average field variance at the held-out source positions

\begin{equation}
\sigma^2_{\textrm{field},i,e}(\lambda) = \E_k\left\{ \E_{j\in S_k}\left[\textrm{Var}\left(g^{\overline{S}_k}_{i,e,\lambda}(\vec{\theta}_j)\right)\right]\right\}.
\end{equation}
Finally, we measure the overall statistical error of estimator $e$ by subtracting the above two error terms
\begin{equation}
\sigma^2_{\textrm{source},i,e}(\lambda) \cong \sigma^2_{\textrm{field-source},i,e}(\lambda) - \sigma^2_{\textrm{field},i,e}(\lambda)
\end{equation}\\

\begin{figure}
\begin{center}
\includegraphics[width=0.9\columnwidth]{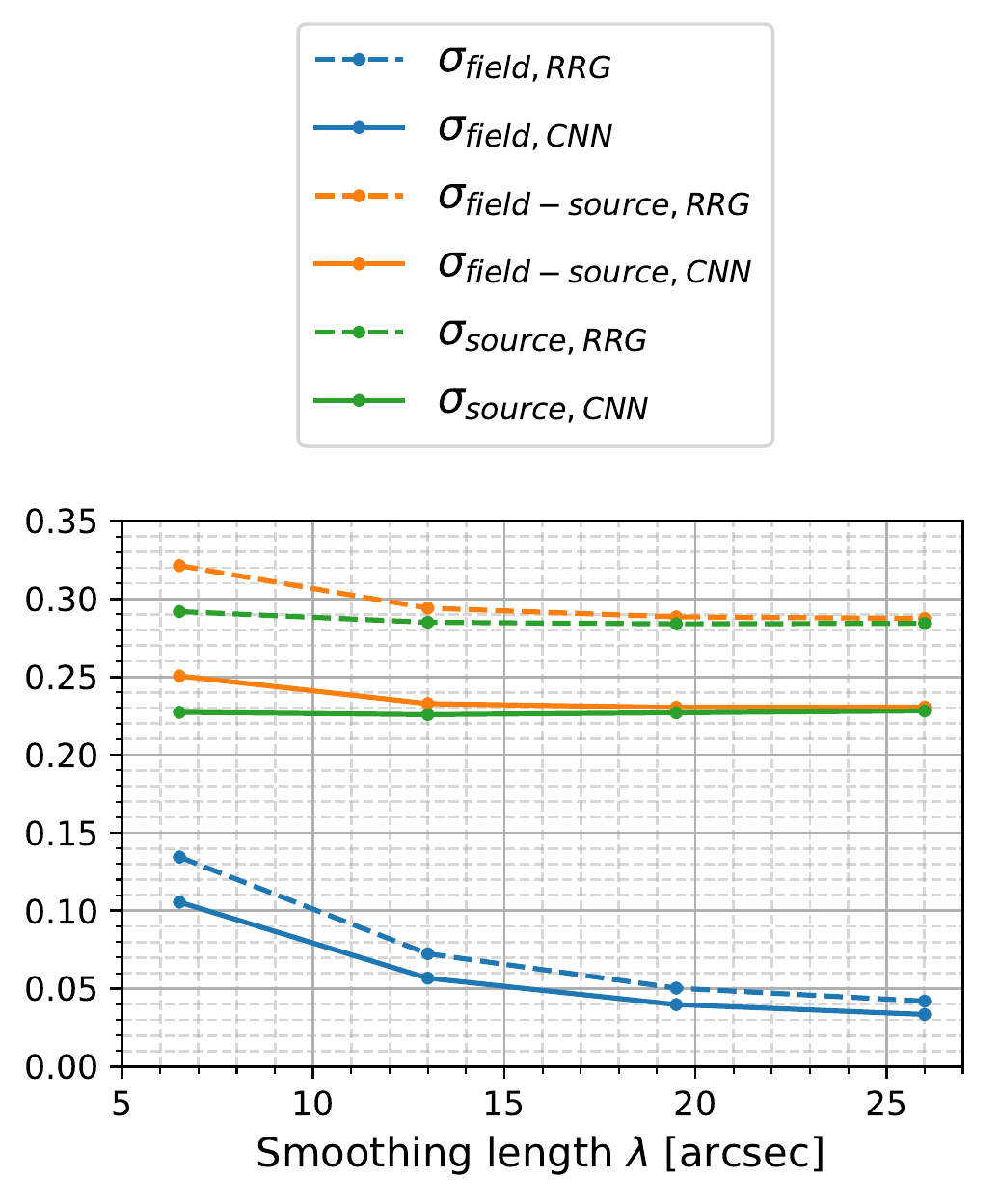}
\end{center}
\caption{Statistical errors of the smooth field estimates ($\sigma_\textrm{field}$), the mean scatter between the field and per source estimates ($\sigma_\textrm{field-source}$) and the statistical errors of the per source estimates ($\sigma_\textrm{source}$). Here we present the average standard deviation of the two shear components as a function of field smoothing length $\lambda$ for the RRG and CNN estimators.}
\label{fig:clash_error_analysis}
\end{figure}

Figure \ref{fig:clash_error_analysis} shows the dependence of these three statistical errors on the smoothing length. We see that both $\sigma_{\textrm{field-source},e}(\lambda)$ and $\sigma_{\textrm{field},e}(\lambda)$ fall with increasing smoothing length $\lambda$. This is qualitatively expected due to the reduced finite-sample-noise in the smooth field estimates at greater values of $\lambda$. The statistical errors of the per source estimates, $\sigma_{\textrm{source},e}(\lambda)$, are approximately constant in $\lambda$. This is consistent with the fact that the per source estimate errors should be statistically independent of the field estimate errors. Overall we find RMS errors of $0.286 \pm 0.003$ for the RRG estimator and $0.227 \pm 0.006$ for the CNN estimator on the CLASH data, a reduction of approximately $26\%$ in statistical errors. The RMS errors we measure for the RRG estimator are in agreement with the shape-noise \cite{leauthaud2007weak} measures using RRG for a population having similar magnitudes. There, shape-noise was estimated using in-the-field galaxies, unaffected by cluster lensing shear. The $\sigma_{\textrm{field}}$ curves in Figure \ref{fig:clash_error_analysis} also illustrate the resulting enhancement in the effective angular resolution of the smooth field estimates at a set level of noise.\\

\label{sec:statistical_errors_clash}

\section{Code, Data and Reproducibility}
\label{sec:reproducibility}

We provide online\footnote{\url{http://github.com/ofersp/wlenet}} the python code and data used in this work. This includes procedures to match and simulate a target background source distribution, the training and testing procedures we used on the simulated data, as well as our weak lensing analysis pipeline. To enable the reader to fully reproduce our work, we release the datasets used to train and test the models described, the specification of the neural network architectures as well as our resulting trained model weights. Regarding the 20 CLASH cluster cores analysed in this work, we publish a set of tables with the updated point source classification described in \S\ref{sec:point_sources}, updated weak lensing shear catalogues (per background source) produced by our CNN model. Additionally we provide the weak lensing fields (per cluster) evaluated in \S\ref{sec:smooth_fields} at various smoothing length scales. Finally, we publish a set of jupyter notebooks\footnote{\url{https://jupyter.org}} \citep{kluyver2016jupyter} with which we performed the analysis and generated the figures included in this paper.

\section{Discussion}
\label{sec:discussion}

We present a machine learning process that allowed us to train a model to perform weak lensing shear estimation directly from simulated examples. This process includes: Simulating galaxy stamps having known shear and an observational distribution matching that of the target stamps, training a \textit{deep convolutional neural network} on the simulated data to produce a model and an assessment of the model performance on both the simulated data and actual galaxy cluster field observations.

This approach is novel in several respects: The model is trained to directly estimate lensing shear and not galaxy shapes, it is not based on hand-crafted features but instead attempts to extract all relevant information from the galaxy observations (in the limits of the model architecture) and it is designed to be invariant to the stamp centroid and background estimates.

We dealt with the following challenges that were specific to our approach: Avoiding simulation artefacts which would otherwise deteriorate model generalization to actual observations, enlarging and augmenting the training data to promote wanted invariance properties at the learning phase and enforcing known discrete symmetries of lensing shear at the inference phase. In the observed CLASH fields, an additional challenge was to perform bias calibration and statistical error analysis where the true per-stamp shears are unknown.

The results of our learnt estimator appear to be consistent with those of commonly used quadrupole moment estimators and show enhanced statistical performance. We find this to be the case both on the simulations and in the real observations where the reduction in RMS errors, compared to the RRG estimator, is at the level of 26\%. This is equivalent to a relative improvement in survey speed of approximately 60\%. This serves as initial indication that what is commonly referred to as the \textit{shape-noise limit} can likely be overcome, at least to some extent, by using the joint statistics of higher order intensity moments of high-redshift galaxy images (above the second order moment statistics tapped by RRG).\\

These results however require additional testing and validation due to the non-transparent nature of the machine learning approach. Although we have made great efforts to rid our simulations of artefacts, these may still exist in our training and test datasets and could potentially be providing the learnt estimator with clues as to what the simulated shears were --- clues that would not be available in real observations. Additional simulations by independent means would allow us to mostly rule out this possibility. Furthermore, studying the error sensitivity (statistical and systematic) of our estimator to additional distributional factors, such as galaxy morphology \citep[using e.g. the Shapelet decomposition of][at increasing order]{refregier2003shapelets}, S/N levels and PSF parameters, is still required. This would enable us to disentangle the various sources of statistical error (e.g. morphology from Poisson noise) and to control potential systematic biases when applying this learnt estimator to different real observing conditions.\\

Using high dimensional statistics to estimate shear can either be done explicitly, by modelling the conditional distribution (conditioned on shear) of post-shear galaxy images --- this is known as a \textit{generative approach} to machine learning, or implicitly, by learning to fit a function that directly maps post-shear images of galaxies to their true shear, in a statistically consistent and efficient way --- known as a \textit{discriminative approach}. This later approach is the one we eventually followed in this work. This choice requires us to have good simulations at our disposal \citep{rowe2015galsim} but allows us to avoid analytically modelling the action of shear on the pre-sheared galaxy image distribution. Training our model in a way that promotes invariance to certain types of transformations is also simplified by the discriminative approach --- achieving this only requires us to simulate these transformations when preparing the training data.\\

We list the following caveats of our approach:
\begin{itemize}
\item
Our estimator does not presently deal with PSF variation. We believe this does not affect the validity of our statistical error analysis, particularly on the simulated test data, but could in principle be leading to local biases of the estimator within the observed field-of-view in the CLASH data --- possible biases which we have not accounted for. We believe this issue can be overcome by training a model to infer shear given both a galaxy stamp and a local estimate of the PSF, such as those made available by the \textsc{TinyTim} PSF model of the \textit{HST} fields \citep{krist201120}. We defer this issue to future work.

\item
Assuming a total galaxy number density of $37\,\,\mathrm{arcmin}^{-2}$ \citep[see][]{chang2013effective} we estimate that in the well resolved CLASH data ${\sim}4\%$ should contain blends within a $32\times 32\,\textrm{px}$ stamp, although the vast majority of these blends would be of all-background galaxies. For all-background blends, a similar shear would affect the whole stamp, as is the case in a similar proportion of the stamps in the COSMOS based simulations. In addition, blends including foreground galaxies typically manifest as constant or nearly constant additive background levels within the $32\times 32\,\textrm{px}$ background stamps. As discussed in \S\ref{sec:sensitivity}, our estimator is expected to be robust to this type of blending. Stamps with blends that include relatively small foreground galaxies may introduce some systematic biases which we have not accounted for. Although we do not expect blending to introduce a significant effect in the CLASH observational regime, simulating these foreground-background blends in the training data could allow us to both train the CNN estimator to be robust to blending issues and to assess its sensitivity to blending. We defer this refinement of our simulation process to future work.

\item
Our simulations are lacking in that we do not match the simulated and observed populations in redshift, but only in magnitude-radius space (see \S\ref{sec:mag_rad_matching}). This choice could in principle be overcome, to a large extent, when deeper observations of in-the-field galaxies become available. This is due to the fact that the higher the S/N of the base dataset used by \textsc{GalSim}, the closer the image rescaling factor ($\alpha$ in \S\ref{sec:sn_tradeoffs}) can be to unity. The availability of deeper observations at the training phase, particularly when provided in the specific band we later wish to measure weak lensing shear in, will also allow us to avoid the band mismatch we opt for in \S\ref{sec:acs_total_images}.

\item
The reliance of our approach on high-fidelity image simulations adds difficulty to its application. This is similarly the case with conventional simulation-based shear calibration techniques and the potential advantage of data-based calibration \citep[e.g.][]{huff2017metacalibration}.

\item
When producing the training data for our learnt estimator we performed a set of transformations, or augmentations, in addition to the simulated shear. The choice of augmentations was not particularly tuned to the CLASH dataset which we later ran our estimator on, but these were chosen after we were exposed to the CLASH data. As previously discussed, testing our model on additional simulations as well as on real galaxy cluster observations are needed to further validate our approach.

\end{itemize}

Our technique was developed and tested in the regime of well-resolved space-based images. This was done with particular aim at galaxy cluster weak lensing measurements. The applicability of this technique to the type of ground-based images used in cosmic shear measurements is currently not known and remains to be studied. It is likely that the lack of visible substructure in the ground-based image regime would lead to little gain for this approach.\\

Finally, we would like to stress the point raised in \S\ref{sec:arch} --- we are currently unaware of any justified strategies to choose an optimal model architecture, model weight optimization scheme and training data augmentation scheme. We would like to encourage the reader to use the code and data released in this work as a basis for future experimentation, hopefully, reaching lower statistical errors and further enhancing resolving power in real galaxy cluster weak lensing observations.

\section{Acknowledgements}
We thank Anton Koekemoer, Rachel Mandelbaum, Elinor Medezinski, Keichi Umetsu, Adi Zitrin and Boaz Nadler for fruitful discussions. E.O.O. is grateful for support by
a grant from the Israeli Ministry of Science, Minerva, BSF, BSF transformative program, and the I-CORE Program of the Planning
and Budgeting Committee and The Israel Science Foundation (ISF) (grant No 1829/12). O.M.S., E.O.O., and Y.W. are grateful for support by
a grant from the ISF. J.M. has received funding from the European Union's Horizon 2020 research and Innovation programme under the Marie Sk\l{}odowska-Curie grant agreement No 664931. We thank the anonymous reviewer whose comments helped us
improve the manuscript.

\appendix
\section{CNN Layers}
\label{sec:layers}
In the architecture of \S\ref{sec:arch}, a \textit{convolutional layer}, denoted $\texttt{Conv}\,(n,m)\times l$, transforms an input tensor $I_{x,y,c}$, having dimensions $n_i\times m_i\times l_i$, to an output tensor $O_{x,y,c}$ having dimensions $n_o\times m_o\times l$
\begin{equation}
O_{x,y,c} = B_c + \sum_{x'=1}^n \sum_{y'=1}^m \sum_{c'=1}^{l_i} F_{x',y',c'}^{(c)} I_{x+x',y+y',c'}\,\,,
\end{equation}
where $w_\mathrm{conv}=(B_d, F_{x,y,c}^{(d)})$ are the bias and filter weight parameters associated with this layer.\\

An \textit{affine layer}, denoted $\texttt{Affine}\,(l)$, transforms an input tensor $I_{k}$ (or a tensor $I_{x,y,c}$ that has been flattened appropriately), having dimension $n_i$, to an output tensor $O_{j}$ having dimension $l$
\begin{equation}
O_{j} = B_j + \sum_{k=1}^{n_i} A_{jk} I_k\,\,,
\end{equation}
where $w_\mathrm{affine}=(B_j, A_{jk})$ are the bias and linear weight parameters of this layer.\\

Convolutional layers, in our architecture, also perform the $\texttt{Stride}(\Delta x, \Delta y)$ operation, which downsamples an input tensor $I_{x,y,c}$ to an output tensor $O_{x,y,c}$ in the following way
\begin{equation}
O_{x,y,c} = I_{x,y,c} \downarrow_{\Delta x, \Delta y}.
\end{equation}

Both convolutional and affine layers, in our architecture, perform the point-wise non-linear operation defined by $O = \texttt{ReLU}(I) = \max(0, I)$.

Finally, a training technique that promotes generalization (enhances the performance of the resulting model on the test set) known as Dropout \citep{srivastava2014dropout} is employed in some of the layers of Table~\ref{tbl:arch}.

\section{Regression attenuation}
\label{sec:regression_attenuation}

Here we show that performing linear least squares regression in the inverse direction, that is, from noisy features to clean signal, produces a biased estimator (a bias known in statistics as \textit{regression attenuation}). Under some model assumptions regarding the joint probability of noise and signal we find a simple correction factor that transforms the inverse regression estimator to the conditionally unbiased linear estimator having least variance.

In this appendix we use $x \in {\mathbb R}$ to denote the signal we wish to estimate and $y \in {\mathbb R^d}$ to denote the $d$-dimensional noisy feature vector which we measure. To simplify the discussion we assume the signal $x$ has a prior distribution having zero mean, $\E\left[x\right] = 0$. The components of reduced shear in \S\ref{sec:train_and_aug} are represented here by $x$ and this zero mean assumption is compatible with how the reduced shear was prepared in the training data there. Our model for how the noisy features are generated is the following
\begin{equation}
\label{eq:noisy_feature_model}
y = sx + \eta,
\end{equation}
where $s \in {\mathbb R^d}$ is some set of constant coefficients and $\eta \in {\mathbb R^d}$ is noise uncorrelated with the signal, $\E[x\eta] = 0$. We also assume the noise has zero mean for every $x$, $\E[\eta|x] = 0$. We found that both of these properties approximately hold in \S\ref{sec:train_and_aug} for the noisy features after a constant mean is subtracted. We now confine our estimator $\hat{x}$ of the signal $x$ to be some linear function of the noisy features $y$,
\begin{equation}
\hat{x} = a^{T} y,
\end{equation}
for some set of constant coefficients $a \in {\mathbb R^d}$. 

For a set of $N$ samples of signals $X = \left(x_1, x_2, \cdots, x_N\right)^T$ and matching features $Y = \left(y_1, y_2, \cdots, y_N\right)^T$ we can write equation \ref{eq:noisy_feature_model} in matrix form as 
\begin{equation}
\label{eq:noisy_feature_model_mform}
Y = Xs^T + W,
\end{equation}
with $W = \left(\eta_1, \eta_2, \cdots, \eta_N\right)^T$. Performing linear-least-squares regression from the noisy features to the clean signal produces the following estimator coefficients
\begin{equation}
a_{\textsc{MMSE}} = \argmin_a L(a) = \argmin_a \sum_{i=1}^N\left(a^T y_i - x_i\right)^2.
\end{equation}
This is the minimum-mean-squared-error (MMSE) estimator which can be shown to have the explicit form 
\begin{equation}
a_{\textsc{MMSE}} = \left(Y^T Y\right)^{-1} Y^T X
\end{equation}
when $Y^T Y$ is invertible. For a sufficiently large number of samples, this can be shown to be 
\begin{eqnarray}
\label{eq:a_mmse}
a_{\textsc{MMSE}} &=& \frac{1}{N} C_y^{-1} Y^T X \\
&=& \frac{1}{N} C_y^{-1} (s X^T + W^T) X \\
&=& C_y^{-1} (\frac{1}{N} s X^T X + \frac{1}{N} W^T X) \\
&=& C_y^{-1} \left\{\mathrm{Var}(x) s + \E\left[x\eta\right]\right\} \\
&=& \mathrm{Var}(x) C_y^{-1} s,
\end{eqnarray}
where $C_y$ is the $d \times d$ covariance matrix of $y$. 

Different estimator coefficients would have resulted if we would have looked for the conditionally unbiased linear estimator --- the $\hat{x}$ for which
\begin{equation}
\forall x\,x = \E[\hat{x} | x] = a^T s x,
\end{equation}
having minimal variance
\begin{equation}
\mathrm{Var}(\hat{x}) = \E\left[\left(\hat{x} - \E[\hat{x}]\right)^2\right] = a^T C_y a.
\end{equation}
These two last requirements lead to the following constrained minimization problem
\begin{equation*}
\begin{aligned}
& \underset{a}{\text{minimize}}
& & a^T C_y a \\
& \text{subject to}
& & a^T s = 1,
\end{aligned}
\end{equation*}
for constant $C_y$ and $s$. Using the method of Lagrange multipliers, this can be shown to have the solution
\begin{equation}
\label{eq:a_star}
a^* = \frac{C_y^{-1}s}{s^T C_y^{-1}s}.
\end{equation}
We see that the ratio of these two sets of estimator coefficients, \ref{eq:a_mmse} and \ref{eq:a_star}, is the following scalar factor
\begin{equation}
\frac{a^*}{a_\textsc{MMSE}} = \frac{\mathrm{Var}(x)^{-1}}{s^T C_y^{-1}s}.
\end{equation}
In the large sample number limit, this ratio is 
\begin{equation}
\frac{a^*}{a_\textsc{MMSE}} = \left[1 - \frac{L(a_\textsc{MMSE})}{\textrm{Var}(x)}\right]^{-1}.
\end{equation}


\bibliographystyle{mnras}
\bibliography{wlenet} 


\bsp	
\label{lastpage}
\end{document}

%% file: clash_clusters.tex
\begin{table}
\caption{Properties of the 20 CLASH clusters included in our analysis. Here $z_{\mathrm{clust}}$ are the cluster redshifts provided by \protect\cite{postman2012cluster}, $A_{\mathrm{eff}}$ is the fully exposed angular area in $\mathrm{{\prime}^{2}}$ (see \S\ref{sec:point_sources}), $N_{\mathrm{wl}}$ is the number of sources in the full weak lensing catalogue of \protect\cite{merten2015clash}, $N_{\mathrm{cut}}$ is the number of sources in our subset catalogue after the additional cuts of \S\ref{sec:point_sources}, $\beta_{\mathrm{eff}}$ is the mean lensing efficiency \citep[see][\S4.3.2 for a definition of lensing efficiency, or \textit{lens strength}]{bartelmann2001weak} of sources in the cut \protect\citep[$\beta_{\mathrm{eff}}$ assumes a flat cosmological model similar to a WMAP7
cosmology as in][with $\Omega_m = 0.27$, $\Omega_\Lambda = 0.73$ and a Hubble constant of $h = 0.7$]{komatsu2011} and $n_{\mathrm{cut}}$ is the number density of the cut sources in $\mathrm{{\prime}^{-2}}$.}
\label{tbl:clash_catalog_wl}
\begin{tabular}{lcccccr}
\hline \hline
Name & $z_{\mathrm{clust}}$ & $A_{\mathrm{eff}}$ & $N_{\mathrm{wl}}$ & $N_{\mathrm{cut}}$ & $\beta_{\mathrm{eff}}$ & $n_{\mathrm{cut}}$ \\
 &  & $\mathrm{{\prime}^{2}}$ &  &  &  & $\mathrm{{\prime}^{-2}}$ \\
\hline
Abell 1423 & 0.213 & 14.4 & 807 & 442 & 0.738 & 30.7 \\
Abell 209 & 0.206 & 14.4 & 832 & 323 & 0.759 & 22.5 \\
Abell 2261 & 0.224 & 17.5 & 725 & 441 & 0.710 & 25.2 \\
Abell 383 & 0.187 & 16.4 & 796 & 574 & 0.767 & 34.9 \\
MACSJ0329-02 & 0.45 & 15.2 & 493 & 359 & 0.593 & 23.6 \\
MACSJ0416-24 & 0.396 & 14.5 & 551 & 326 & 0.632 & 22.5 \\
MACSJ0429-02 & 0.399 & 14.4 & 654 & 393 & 0.594 & 27.2 \\
MACSJ0647+70 & 0.584 & 17.1 & 773 & 384 & 0.484 & 22.5 \\
MACSJ1115+01 & 0.352 & 14.4 & 491 & 331 & 0.645 & 23.0 \\
MACSJ1149+22 & 0.544 & 13.7 & 844 & 642 & 0.460 & 46.9 \\
MACSJ1206-08 & 0.44 & 14.8 & 581 & 345 & 0.600 & 23.4 \\
MACSJ1311-03 & 0.494 & 14.4 & 447 & 288 & 0.503 & 20.0 \\
MACSJ1720+35 & 0.391 & 16.6 & 635 & 331 & 0.626 & 19.9 \\
MACSJ1931-26 & 0.352 & 14.3 & 709 & 515 & 0.595 & 36.0 \\
MACSJ2129-07 & 0.57 & 14.3 & 853 & 528 & 0.519 & 37.0 \\
MS 2137.3-2353 & 0.313 & 14.4 & 801 & 399 & 0.695 & 27.6 \\
RXJ1347-1145 & 0.451 & 14.3 & 633 & 339 & 0.577 & 23.8 \\
RXJ1532.9+3021 & 0.345 & 16.2 & 508 & 310 & 0.648 & 19.2 \\
RXJ2129+0005 & 0.234 & 14.4 & 624 & 311 & 0.701 & 21.6 \\
RXJ2248-4431 & 0.348 & 15.1 & 598 & 366 & 0.651 & 24.2 \\
\hline
\end{tabular}
\end{table}

%% file: arch.tex
\begin{table}
\caption{Layers of the CNN architecture used in this work, computed left-to-right and top-to-bottom. Overall this architecture has 107,866 parameters. See Appendix~\ref{sec:layers} for a definition of the layers and operations mentioned here.}
\label{tbl:arch}
\begin{tabular}{lcr}
\hline \hline
Layer operations & Output Shape & Params. \\
\hline
\texttt{Input} & $32\times 32\times 1$ & - \\
\texttt{Conv} $(5,5)\times 64$, \texttt{Stride} $(3,3)$, \texttt{ReLU} & $10\times 10\times 64$ & 1,664 \\
\texttt{Conv} $(1,1)\times 40$, \texttt{ReLU} & $10\times 10\times 40$ & 2,600 \\
\texttt{Conv} $(1,1)\times 10$, \texttt{ReLU} & $10\times 10\times 10$ & 410 \\
\texttt{Affine} $(100)$, \texttt{ReLU}, \texttt{Dropout} & 100 & 100,100 \\
\texttt{Affine} $(30)$, \texttt{ReLU}, \texttt{Dropout} & 30 & 3,030 \\
\texttt{Affine} $(2)$ & 2 & 62 \\
\hline
\end{tabular}
\end{table}

%% file: sim_datasets.tex
\begin{table}
\caption{Simulated datasets generated by augmenting mutually exclusive training and test random subsets of the original COSMOS $M_{\textrm{F814W}} < 23.5,\,25.2$ sources, as discussed in \S\ref{sec:train_and_aug}.}
\label{tbl:sim_datasets}
\begin{tabularx}{\columnwidth}{lccccc}
\hline \hline
&
\multicolumn{2}{c}{Original} & \multicolumn{1}{c}{Aug. (i-iv)} & \multicolumn{2}{c}{Aug. (iv)} \\
\cmidrule(r){2-3} \cmidrule(r){4-4} \cmidrule(r{1em}){5-6}
& Train & Test & Train & Train & Test \\
\hline
COSMOS 25.2 & $65\,756$ & $21\,994$ & $1.5$M & $308$K & $308$K \\
COSMOS 23.5 & $22\,423$ & - & $750$K & - & - \\
\hline
\end{tabularx}
\end{table}

%% file: perf_overall.tex
\begin{table}
\caption{Calibration parameters and RMS errors evaluated on the full CLASH matched test set. Here $c_i$ are the additive biases, $m_i$ are the multiplicative biases and $\sigma_i$ are the RMS errors of the bias corrected estimates $\hat{g}_{e,i}$ of estimators $e \in \{\textsc{cnn},\textsc{rrg}\}$.}
\label{tbl:perf_overall}
\begin{tabular}{lcccccr}
\hline \hline
 & $c_1$ & $m_1$ & $c_2$ & $m_2$ & $\sigma_1$ & $\sigma_2$ \\
\hline
RRG& 0.001 & 0.119 & 0.000 & 0.121 & $0.257(1)$ & $0.259(1)$ \\
CNN& 0.000 & 0.005 & 0.000 & 0.005 & $0.225(1)$ & $0.224(1)$ \\
\hline
\end{tabular}
\end{table}

%% file: clash_biases.tex
\begin{table}
\caption{Calibration parameters for the CNN estimator on the CLASH dataset. Here $c_i$ are the additive biases and $m_i$ are the multiplicative biases that transform (according to Equation \ref{eq:bias_correction}) the simulation calibrated CNN estimator such that on the CLASH dataset it is unbiased relative to the RRG catalogue.}
\label{tbl:clash_biases}
\begin{tabular}{cccccr}
\hline \hline
 $c_1$ & $m_1$ & $c_2$ & $m_2$ \\
\hline
$0.000\pm0.002$ & $0.048\pm0.024$ & $0.005\pm0.002$ & $0.057\pm0.024$ \\
\hline
\end{tabular}
\end{table}